\documentstyle[12pt]{article}
\begin{document}

\begin{center}
{\Large Three-body treatment of the penetration through the Coulomb field 
of a two-fragment nucleus} \\ [4mm] 
{\large V.F. Kharchenko}$\footnote[1]{\mbox{E-mail: 
vkharchenko@bitp.kiev.ua}}$  
{\large and A.V. Kharchenko} \\[1mm]
{\small\it Bogolyubov Institute for Theoretical Physics, 
National Academy of Sciences of Ukraine, UA - 03143, Kyiv, Ukraine} \\ 
[1cm]

\noindent\rule{\textwidth}{0.1mm}
\end{center} 

\noindent {\bf Abstract}\\ [1mm]

On the basis of the Faddeev integral equations method and the Watson-
Feshbach concept of the effective (optical) interaction potential, the 
first fully consistent three-body approach to the description of the 
penetration of a charged particle through the Coulomb field of a two-particle 
bound complex (composed of one charged and one neutral particles) has been 
developed. A general formalism has been elaborated and on its basis, 
to a first approximation in the Sommerfeld parameter, the influence of 
the nuclear structure on the probability of the penetration of a charged 
particle (the muon, the pion, the kaon and the proton) through the 
Gamow barrier of a two-fragment nucleus (the deuteron and the two lightest 
lambda hypernuclei, $^3_{\Lambda}\mbox{H}$ and $^5_{\Lambda}\mbox{He}$) 
has been calculated and studied. \\

\noindent {\it PACS:} 03.65.Nk; 21.45.+v; 21.80.+a \\
{\small \it Keywords:} {\small Three-body formalism; Coulomb interaction;
Gamow barrier of two-fragment nucleus; Penetration factor; Deuteron;
Lambda hypernuclei $^3_{\Lambda}\mbox{H}$ and $^5_{\Lambda}\mbox{He}$} \\ [3mm]

\pagebreak
 
\noindent {\bf 1. Introduction} \\ 

The real nuclear systems, as a rule, contain charged particles. 
The interplay of the finite-range nuclear and long-range Coulomb forces
is one of the fundamental peculiarities of the nuclear systems that is
responsible for occuring specific phenomena.

The Gamow theory of the radioactive decay of heavy nuclei [1, 2] was
among first advantageous applications of the quantum mechanics to the
description of atomic nuclei. According to the theory, the probability
of the alpha-decay of a nucleus is essentially determined by the 
probability of the penetration of the $\alpha$-particle from the inside of
the nucleus where the nuclear force dominates to the outside through
the Coulomb barrier of the nucleus. Presence of the Coulomb interaction
between particles can also effect essentially on characteristics of
nuclear processes involving charged particles, most pronouncedly at
low energies.

Of particular interest is the investigation of the effects of the
Coulomb interaction for the simplest nuclear systems at the 
microscopic level --- on the basis of the rigorous few-body equations.
The application of the few-body approach to the problem of scattering a 
charged particle by the Coulomb field of a composite bound system
containing  charged particles makes it possible to reveal new regularities
for processes initiated by the collision. In particular, this approach
enables one to investigate the influence of the structure of the 
scattering complex on the magnitude of the penetration factor for the 
Coulomb field of the complex. 

It is known [3] that the Faddeev integral equations for three particles
[4] with the Coulomb interaction between the particles are non-Fredholm 
at energies above the threshold of the elastic scattering of a particle on 
a two-particle bound system and therefore they must be rebuilt. For the 
problem of the scattering of two charged particles, a convenient method of 
solving the non-Fredholm Lippmann-Schwinger integral equation with the 
Coulomb potential that is based on using the limiting transition from the 
equation with the screened Coulomb potential has been elaborated by 
Gorshkov [5]. Because of the non-Fredholm character of the kernel of 
the two-particle integral equation with the "pure" Coulomb potential 
(in three-dimensional formulation), the solution of the scattering problem 
for two charged particles is known not to be the immediate limiting case 
of the solution of the problem with the screened Coulomb potential when 
the screening radius $R_s$ is going to infinity. The Gorshkov recipe of 
the transition consists in extraction explicitly from the wave function 
of the system with the screened Coulomb potential of the known exponential 
Coulomb factor, singular at $R_s \rightarrow \infty $. The generalization 
of the Gorshkov's recipe for the problem of the scattering of a charged 
particle on a bound system consisting of two charged particles has 
been proposed by Vesselova [6].

The appropriate regularized integral equations of the proton-deuteron
scattering problem were obtained in the paper [7]. In the frame of the
quasi-particle AGS formalism [8] the Vesselova method has been also
used in the papers [9,10]. A further numerical calculations have shown
the benefit from a correct accounting of the Coulomb effects (especially
in the "inner" range --- in the range of the deuteron system) using the 
integral equations method in the momentum space [9,11--13] as well as 
the integro-differential equations method in the configuration space [14].

As a result of the analytical study of the effective interaction 
between a charged particle and two-particle complex on the basis of 
the Faddeev integral equations, it has been established that the full 
concellation of the long-range terms proportional to $(e_1e_2)^2\mid{\bf 
p - p'}\mid^{-1}$, $(e_1e_2)^3\ln (\mid{\bf p - p'}\mid\rho_0)$ and 
$(e_1e_2)^4\mid{\bf p - p'}\mid$ occur when the transfer momentum tends 
to zero ($\mid{\bf p - p'}\mid\rightarrow 0$) and the behaviour 
of the polarization potential at asymptotically large [15--18] and 
intermediate [19--20] distances has been determined in an explicit form. 

The advance in elaboration of the description of the three-particle system 
containing charged particles makes it possible to formulate and investigate 
the problem on the penetration of a charged particle through the Coulomb 
field of a two-particle bound system on the mathematically rigorous level. 

In this paper we develop the first fully consistent three-body approach
to the description of the penetration of a charged particle through the
Coulomb field of a two-fragment bound nucleus. In Section 2 we begin with 
formulation of this problem as a three-body problem on the Coulomb 
scattering of a charged particle by a two-body bound system composed of one 
charged and one neutral particles. Using the Watson-Feshbach concept of the 
effective potential we reduce in a rigorous way the three-body problem
to a definite two-body scattering problem. In Section 3 we elaborate a 
three-body formalism for the description of the penetration of a charged 
particle through the Coulomb field of the two-body nuclear complex. 
We derive a general formula for the relative difference between the 
penetration factor for the Gamow barrier of a two-fragment nucleus and 
the penetration factor for the Coulomb field of the corresponding point 
charge and study its behaviour on simplifying assumptions. In Section 4 
we calculate the effect of influence of the structure of two-fragment 
nuclei (the deuteron and the lambda-hypernuclei $^3_{\Lambda}\mbox{H}$ 
and $^5_{\Lambda}\mbox{He}$) on the probabilities of penetration of 
incident charged particles (the meson, the pion, the kaon and the proton) 
through the Gamow barriers of the nuclei. We summarize our results in 
Section 5. \\ [1mm]

\noindent {\bf 2. Basic equations for the scattering of a charged 
particle by the Coulomb field of a two-particle complex} \\ 

Let us consider the three-particle problem on the scattering of a 
charged particle 1 by a bound complex that consists of two particles --- 
a charged particle 2 and a neutral particle 3. (The particle $i$ is 
characterized by the mass  $m_i$ and the charge $e_i$ .) In this paper we 
restrict our consideration to a three-particle system with two pair
interaction potentials --- a short-range (nuclear) potential describing 
the interaction between the particles  2 and 3, $\;v_{23}^ N\;$, that 
provides formation a bound two-particle complex, and the long-range 
Coulomb potential of the interaction between the particles 1 and 2, 
$\;v_{12}^C\;$.  We regard that the particles 1 and 3 do not interact 
at all, $\;v_{31} = 0\;$.  The operator of the potential interaction 
energy for the considered system of three particles has the form 
\begin{equation} 
V = v^C_{12} + v^N_{23} .  
\end{equation}

Note, that the description of the three particle system with the
potential (1) is the simplest three-particle Coulomb problem. This
problem can have an independent physical interest or arise up as an
auxiliary problem when finding the integral kernels of the modified
Faddeev equations for the system containing charged particles [21]. \\ 

\noindent { \it  2.1 Three-particle equations with the screened
Coulomb interaction} \\ 

To derive the integral equations for the scattering of a charged particle
on a two-particle complex consisting of a charged and neutral particles, we 
start from the interaction operator for the three-particle system 
$\bar{V}$ that contains the screened Coulomb pair potential $\bar{v}_{12}^C$,
\begin{equation}
\bar{V} = \bar{v}_{12}^C + v_{23}^N\;,\quad \bar{v}_{12}^C = v_{12}^C 
\cdot s_{12}, 
\end{equation} 
where $v_{12}^C$ is the ordinary  "pure" Coulomb potential, 
$\; v_{12}^C(r_{12}) = e_1 e_2/r_{12}\;$, and $\;s_{12}$ is a screening 
function that can be chosen in the exponential form: $s_{12}(r_{12},R_s) = 
\exp (- r_{12}/R_{s})\;$,$\; R_s$ is a parameter that describes the 
screening distance. Here and further we mark all quantities, which 
correspond to the screened Coulomb potential, by bar.

Splitting (correspondingly to the form of the potential operator (2))
the total wave function of the three-particle system $\bar{\Psi}$
into two Faddeev's components,
\begin{equation}
\bar{\Psi}=\bar{\Psi}^{(23)}+\bar{\Psi}^{(12)},
\end{equation}
for the components we obtain the system of the coupled integral equations:  
\begin{eqnarray}
 \bar{\Psi}^{(23)}= & \Phi_{23,1}\; + & G_0(E) T_{23}(E)\bar{\Psi}^{(12)}, 
 \nonumber \\
 \bar{\Psi}^{(12)}= &            & G_0(E) \bar{T}_{12}^C(E)\bar{\Psi}^{(23)}.
 \label{line2}
 \end{eqnarray}
Here, the operator $G_0(E)=(E - h_0^{23}- h_0^1 + i0)^{-1}$ is the free 
three-particle propagator, ($h_0^{23}$ and $h_0^1$ are the operators of 
the kinetic energy of the relative motion of the particles 2 and 3 and 
that of the particle 1 and the centre of mass of the particles 2 
and 3),  $E$ is the total energy of the relative motion in the system 
of three particles, $\bar{T}_{12}^C(E)\;$ and $\;T_{23}(E)$ are the 
transition operators that correspond to the pair interactions 
$\bar{v}_{12}^C\;$ and $\;v_{23}$, respectively, and satisfy to the 
Lippmann-Schwinger equations with the free three-particle propagator 
$G_0(E)$, 
\begin{eqnarray} 
\bar{T}_{12}^C(E) = \bar{v}_{12}^C + \bar{v}_{12}^C 
G_{0}(E) \bar{T}_{12}^C(E), \nonumber \\ [3mm] T_{23}(E) = v_{23} + v_{23} 
G_{0}(E) T_{23}(E) . \label{line2} 
\end{eqnarray} 
The free term in Eq. (4) is the product of the wave function of the bound 
state of the two-particle complex $\psi_0$ and the plane wave of the 
relative motion of the incident particle 1 
and the centre of mass of two particles of the complex with the momentum 
${\bf p}_0$,$\quad\varphi_{{\bf p}_0}=\mid{\bf p}_0\rangle$, 
\begin{equation} 
\Phi_{23,1} = \psi_{0} \cdot \varphi_{{\bf p}_0}.
\end{equation}

As variables that describe the three-particle system we shall use the
relative Jacoby momenta:
\begin{equation}
{\bf k}_{ij} = (m_j {\bf k}_i - m_i 
{\bf k}_j)/m_{ij},\hspace{12mm} {\bf p}_{k} = \left[ m_k 
\left({\bf k}_i + {\bf k}_j\right) - m_{ij}{\bf k}_k\right]/M, 
\end{equation} 
where ${\bf k}_i$ is the momentum of the particle $i$ ,$\; m_{ij} = 
m_i + m_j\;, \hspace{5mm} M = m_1 + m_2 + m_3,\hspace{5mm} ijk = 123, 231, 
312 $. In the configuration space the relative coordinates
\begin{center} 
${\bf r}_{ij} = {\bf r}_i - {\bf r}_j  , \hspace{1.2cm} 
\mbox{\boldmath $\rho$}_{k} = \left( m_i {\bf r}_i + m_j{\bf 
r}_j\right)/m_{ij} - {\bf r}_k $, 
\end{center} 
where ${\bf r}_i$  is the radius-vector of the particle $i$, correspond to 
the relative momenta (7). Below we shall use the short-cut definitions of the 
momentum variables$\;\; {\bf k} \equiv {\bf k}_{23}, \hspace{5mm}
{\bf p} \equiv {\bf p}_1 \hspace{5mm}$ and the coordinates$\;\; {\bf r} 
\equiv {\bf r}_{23}, \hspace{5mm} \mbox{\boldmath $\rho$} \equiv 
\mbox{\boldmath $\rho$}_1$.

For the problem on the scattering of a particle on a two-particle complex
the total energy of the relative motion in the three-particle system is
equal to 
\begin{equation}
E = \epsilon - b ,
\end{equation}
where $\epsilon = p_0^2/2\mu_1$ is the energy of the relative motion of 
the particle 1 and the centre of mass of the complex of the particles 2 and 3,
$b = \kappa^2/2\mu_{23}$ is the binding energy of the two-particle complex,
$\mu_{ij} = m_i m_j/m_{ij}$ \hspace{3mm}÷\hspace{3mm} and
$\mu_k = m_k m_{ij}/M$ are the reduced masses of two particles $i$ and $j$ 
and of the particle $k$ and the centre of mass of the pair of particles, 
$i$ and $j$, respectively.

For simplicity sake the interaction between particles 2 and 3 that form
a bound complex will be described by the separable potential:
\begin{equation}
v_{23}^N = - \lambda \mid u \rangle \langle u \mid .
\end{equation}
To the potential (9) there corresponds the two-particle transition
operator  
\begin{equation}
t_{23}^N(\varepsilon) = \mid u \rangle \tau (\varepsilon) \langle u \mid ,
\end{equation}
where
\begin{equation}
\tau (\varepsilon) = \frac{S(\varepsilon)}{\varepsilon + b}\;,\;\; 
S(\varepsilon) = \langle u \mid g_0^{23}(-b)g_0^{23}(\varepsilon) \mid u 
\rangle^{-1}\;,\;\;\nonumber \\ g_0^{23}(\varepsilon) = 
(\varepsilon - h_0^{23} + i0)^{-1},   \label{line2} 
\end{equation} 
$\varepsilon$  is the energy of the relative motion of the particles 2 and 3.
The wave function of the bound $S$-state of the two particles has the form
\begin{equation} 
\mid \psi_0 \rangle = - g_0^{23}(-b) \mid u \rangle\;,\;\;\; 
\psi_0\left({\bf k}\right) = u(k) (\frac{k^2}{2\mu_{23}} + b)^{-1}\;,\;\;\; 
\langle \psi_0 \mid \psi_0 \rangle = 1, 
\end{equation} 
in this case the occurence of the bound state of two particles implies the
fulfilment of the condition 
\begin{equation} 
\frac{1}{\lambda} + \langle u \mid g_0^{23}(-b) \mid u \rangle = 0.  
\end{equation}

In the case of the potential (9) the transition operator $T_{23}(E)$ 
in the equation of the set (4) that is determined by the Lippmann-Schwinger 
equation (5) with the three-particle free propagator $G_0(E)$, takes 
the form
\begin{equation}
T_{23}(E) = \mid u \rangle \tau(E - h_0^1) \langle u \mid\;,\quad\quad
\tau (E-h_0^1) = S(E-h_0^1) g_0^1(E).
\end{equation}
Note, that the expression (14) for $T_{23}(E)$  contains a diagonal operator
in the functional space of a relative momentum of the particle 1 and the 
centre of mass of the particles 2 and 3. 

By applying the interaction model (9) the formal solution of the set of 
equations (4) can be written as
\begin{eqnarray} 
\bar{\Psi}^{(23)} & = & - G_0(E) \mid u \rangle 
\left[1 + \tau (E- h_0^1) \bar{X}(\epsilon)\right] 
\varphi_{{\bf p}_0}, \nonumber \\ 
\bar{\Psi}^{(12)} & = & - G_0(E)\bar{T}_{12}^C(E)G_0(E)\mid u 
\rangle \left[ 1 + \tau (E- h_0^1) \bar{X}(\epsilon)\right] 
\varphi_{{\bf p}_0}, \label{line2} 
\end{eqnarray} 
where the operator $\bar{X}(\epsilon)$ satisfies the integral equation
\begin{equation} 
\begin{array}{rl} 
\bar{X}(\epsilon) =& \bar{U}(\epsilon) + \bar{U}(\epsilon) 
\tau (\epsilon - b - h_0^1) \bar{X}(\epsilon) , \nonumber \\
& \bar{U}(\epsilon) = \langle u \mid G_0(\epsilon - b) \bar{T}_{12}^C
(\epsilon - b) G_0(\epsilon - b) \mid u \rangle . \label{line2} 
\end{array}
\end{equation}

Taking into account Eq. (15) we write the complete wave function (3) 
in the form
\begin{equation}
\bar{\Psi} = - \left[ 1 + G_0(E) \bar{T}_{12}^C(E)\right] G_0(E)
\mid u \rangle \left[ 1 + \tau \left( E- h_0^1\right) 
\bar{X}(\epsilon)\right] \varphi_{{\bf p}_0}.  
\end{equation}

The operators $\;\bar{X}(\epsilon)\;$, $\;\bar{U}(\epsilon)\;$ and 
$\; \tau (\epsilon - b - h_0^1)\;$ in Eqs. (15) and (16) act in the 
space of the functions of the variables that describe only the relative 
motion of the particle 1 and the centre of mass of the particles 2 
and 3 --- the momentum vector ${\bf p}_1$ or the radius-vector 
$\mbox{\boldmath $\rho$}_1$ . With the formulae (15) - (16), the 
three-body problem under study is reduced to a definite two-body 
problem on the motion of the particle 1 relative to the centre of mass 
of the particles 2 and 3. The latter is formulated in the form of the 
integral equation (16). The attained simplification of the three-body 
problem, feasible because of using a specific model of the interaction  
between the particles of the bound complex (9), is rigorous and can be 
generalized to other models of the finite-range interaction.  \\ 

\noindent { \it  2.2 Equation for the scattering wave function of a
reduced two-body problem} \\ 

Instead of the operator $\bar{X}(\epsilon)$ (matrix elements of which
are contained in the components of the wave function (15) with the initial
momentum on the energy shell $\epsilon$) it is convenient to use the 
relative wave function  
\begin{equation} 
\bar{\chi}_{{\bf p}_0} = \left[ 1 + 
g_0^1(\epsilon) \bar{X}(\epsilon) \right] \varphi_{{\bf p}_0}.  
\end{equation} 
The integral equation for the function $\bar{\chi}_{{\bf p}_0}$
that follows from Eqs. (16) and (18) has the form 
\begin{equation} 
\bar{\chi}_{{\bf p}_0} = \varphi_{{\bf p}_0} + 
g_0^1(\epsilon) \bar{U}(\epsilon)
S \left(\epsilon - b - h_0^1\right) \bar{\chi}_{{\bf p}_0}. 
\end{equation}
With the use of the function $\bar{\chi}_{{\bf p}_0}$ and the relation
\begin{equation}
\left[ 1 + \tau\left(\epsilon - b - h_0^1\right) \bar{X}(\epsilon)\right] 
\varphi_{{\bf p}_0} = S\left(\epsilon - b - h_0^1\right)
\bar{\chi}_{{\bf p}_0} 
\end{equation} 
the expressions for the Faddeev's components (15) and the complete wave 
functions (17) take the form  
\begin{eqnarray} 
\bar{\Psi}^{(23)} & = & - G_0(E) \mid u \rangle S\left(\epsilon 
- b - h_0^1\right) \bar{\chi}_{{\bf p}_0}, \nonumber \\ 
\bar{\Psi}^{(12)} & = & - G_0(E) \bar{T}_{12}^C(E) G_0(E) \mid u \rangle 
S\left(\epsilon - b - h_0^1\right)\bar{\chi}_{{\bf p}_0} \label{line2} 
\end{eqnarray} 
and
\begin{equation} 
\bar{\Psi} = - \left[ 1 + G_0(E) 
\bar{T}_{12}^C(E)\right] G_0(E) \mid u \rangle S\left(\epsilon - b - 
h_0^1\right) \bar{\chi}_ {{\bf p}_0}.  
\end{equation}

The scattering wave function for the effective two-body problem on the
motion of the charged particle 1 relative to the centre of mass of the
particles of the complex, $\bar{\psi}_{{\bf p}_0}^{eff}(1)$, 
we determine following the method of introducing the effective interaction
between a particle and a complex by Francis-Watson [22] and Feshbach
[23] (see also Refs. [24] and [25]) as a coefficient function before the wave 
function of the bound state of the two-body complex in the expansion of the 
complete three-particle wave function in the set of the two-particle wave
functions of the relative motion of the interacting particles 2 and 3, 
\begin{equation} 
\bar{\psi}_{{\bf p}_0}^{eff}(1) \equiv \langle \psi_0(23) \mid 
\bar{\Psi}_{{\bf p}_0}(23,1) \rangle , 
\end{equation} 
In the momentum space of variables the formula (23) has the form
\begin{equation} 
\bar{\psi}_{{\bf p}_0}^{eff}({\bf p}) = \int \frac{d\mbox{\bf k}}{(2\pi)^3} 
\psi_0^*(\mbox{\bf k}) \bar{\Psi}_{{\bf p}_0} (\mbox{\bf k},\mbox{\bf p}) .  
\end{equation}

Taking into account the explicit form of the total wave function (22) in
the expression (23), we obtain the relation between the effective wave 
function $\bar{\psi}_{{\bf p}_0}^{eff}$ and the above-introduced function
$\bar{\chi}_{{\bf p}_0}$, 
\begin{equation} 
\bar{\psi}_{{\bf p}_0}^{eff} = \left[ 1 + \bar{Z}(\epsilon) S\left(\epsilon 
- b - h_0^1\right)\right] \bar{\chi}_{{\bf p}_0}, 
\end{equation} 
where 
\begin{equation} 
\bar{Z}(\epsilon) = - \langle \psi_0 \mid G_0(\epsilon - b) 
\bar{T}_{12}^C (\epsilon - b) G_0(\epsilon - b) \mid u \rangle .  
\end{equation} 
The expression (26) can be written as
\begin{equation} 
\bar{Z}(\epsilon) = g_0(\epsilon)[\bar{W}(\epsilon) - \bar{U}(\epsilon)] ,
\end{equation}
where
\begin{equation}
\bar{W}(\epsilon) =  \langle u \mid g_0^{23}(- b) \bar{T}_{12}^C
(\epsilon - b) G_0(\epsilon - b) \mid u \rangle ,  
\end{equation}
and the operator $\bar{U}(\epsilon)$ is determined by the expression (16). 
The integration in the expressions for $\bar{Z}(\epsilon)$ and 
$\bar{W}(\epsilon)$, as in the expression (16) for $\bar{U}(\epsilon)$, 
is taken only in the space of variables of the relative motion of the
particles inside the complex.  

Introducing the effective transition operator  $\bar{X}_{eff}(\epsilon)$
that corresponds to the effective wave function 
\begin{equation}
\bar{\psi}_{{\bf p}_0}^{eff} =
\left[ 1 + g_0^1(\epsilon) \bar{X}_{eff}(\epsilon)\right] 
{\varphi}_{{\bf p}_0}, 
\end{equation}
and taking into account the relations (25), (27) ÷ (18), we obtain the
formula expressing the transition operator $\bar{X}_{eff}(\epsilon)$ 
through the operators $\bar{X}_(\epsilon)$ and $\bar{W}(\epsilon)$, 
\begin{equation} 
\bar{X}_{eff}(\epsilon) = \bar{W}(\epsilon) + 
\bar{W}(\epsilon) \tau\left(\epsilon - b - h_0^1\right) 
\bar{X}(\epsilon) .
\end{equation}
\\ [3mm]
 
\noindent {\it  2.3 The passage to the unscreened Coulomb interaction} \\

The removal of the screening of the Coulomb interaction between charged 
particles (the passage to the unlimitedly great screening distance,  
$R_s \rightarrow \infty$) in Eq. (19), that determines the
scattering wave function of the reduced two-particle problem and the
effective wave function (24), we perform by applying the known 
Gorshkov-Vesselova recipe [5,6]. Notice that the quantities describing 
the system with the unscreened Coulomb interaction we denote by the same 
letters but without overscribed bar. 

The kernels of the operators $\bar{U}, \bar{W}$ and $\bar{Z}$ in the integral 
equations (19) and (30) and in the expression for the effective wave function 
(25), which are determined by the formulae (16),(26) and (28), has the form
\begin{eqnarray} 
\bar{U}\left({\bf p},{\bf p}'; \epsilon\right) & = & \int 
\frac{d{\bf k}}{(2\pi)^3} \frac{u({\bf k}) \langle{\bf k}_{12}\mid 
\bar{t}_{12}^C\left(\epsilon - b - \frac{p_3^2}{2\mu_3}\right) \mid 
{\bf k}_{12}^{\prime} \rangle u({\bf k}-2{\bf Q})}
{\left [\frac{k^2}{2\mu_{23}} + b - \epsilon_p\right]\left[\frac{( {\bf k}
-2{\bf Q})^2}{2\mu_{23}} + b - \epsilon_{p'}\right]}\;, \nonumber \\ [5mm] 
\bar{W}\left({\bf p},{\bf p}'; \epsilon\right) & 
= & \int \frac{d{\bf k}}{(2\pi)^3} \frac{u({\bf k}) 
\langle{\bf k}_{12}\mid \bar{t}_{12}^C\left(\epsilon - b 
- \frac{p_3^2}{2\mu_3}\right) \mid {\bf k}_{12}^{\prime}  
\rangle u({\bf k}-2{\bf Q})}{\left[\frac{k^2}{2\mu_{23}} 
+ b \right]\left[\frac{({\bf k}-2{\bf Q})^2}{2\mu_{23}} + b - 
\epsilon_{p'}\right]}\;, \label{line2} \\ [5mm] 
\bar{Z}\left({\bf p},{\bf p}'; \epsilon\right) & 
= - & \int \frac{d{\bf k}}{(2\pi)^3} \frac{u({\bf k}) 
\langle{\bf k}_{12}\mid \bar{t}_{12}^C\left(\epsilon - b 
- \frac{p_3^2}{2\mu_3}\right) \mid {\bf k}_{12}^{\prime}  
\rangle u({\bf k}-2{\bf Q})}{\left[ \frac{k^2}{2\mu_{23}}+b\right]
\left[\frac{k^2}{2\mu_{23}} + b - \epsilon_p\right]\left[\frac{(
{\bf k}-2{\bf Q})^2}{2\mu_{23}} + b - 
\epsilon_{p'}\right]}\;, \nonumber  
\end{eqnarray}
where for simplicity sake the following designations are used:
\begin{eqnarray*}
&{\bf k}_{12} = - \frac{m_1}{m_{12}} {\bf p}_{3}  
-{\bf p},\hspace{5mm}{\bf k}_{12}^{\prime}  
= - \frac{m_1}{m_{12}} {\bf p}_{3} -{\bf p}^{\prime},\hspace{5mm}
{\bf p}_{3}={\bf k}- \frac{m_3}{m_{23}} {\bf p},& \\ [2mm]
&{\bf Q} = \frac{m_3}{2m_{23}} ({\bf p}-{\bf p}^{\prime}), 
\hspace*{5mm}\epsilon_p = \epsilon - \frac{p^2}{2\mu_1} + i0.&   
\end{eqnarray*}

In the limiting case of the unscreened ("pure") Coulomb interaction
$(R_s \rightarrow \infty: \bar{v}_{12}^C \rightarrow v_{12}^C)$
the two-particle Coulomb transition matrix $\bar{t}_{12}^C \rightarrow 
t_{12}^C)$ in the integrable expressions (31) generates long-ranged 
(Coulomb and polarization) interactions between the charged incident
particle and the centre of mass of complex. The most long-ranged part of
the kernels of the operators $\left\{U,W,Z\right\} = \lim_{R_s 
\rightarrow \infty}\left\{\bar{U}, \bar{W}, \bar{Z}\right\}$ 
is formed in Eq. (31) from the Born term of the two-particle Coulomb 
transition matrix. The corresponding matrix element being local depends on  
${\bf k}_{12}^{\prime}-{\bf k}_{12} = {\bf p}-{\bf p}^{\prime},$ and
hence is a function of only variables characterizing the relative motion of
the particle 1 and the centre of mass of two other particles, which has
the form of the screened Coulomb potential of interaction between the
point charges of the particle 1 ($e_1$) and the centre of mass of the
two-body complex ($e_2$), $\bar{V}^C({\bf p} -{\bf p}^{\prime})$, 
\begin{equation} 
\langle {\bf k}_{12} \mid \bar{t}_{12}^C (\epsilon - b - 
\frac{p_3^2}{2\mu_{3}}) \mid {\bf k}_{12}^{\prime} \rangle_{Born} = 
\bar{v}_{12}^C({\bf k}_{12} - {\bf k}_{12}^{\prime}) = \frac{4\pi e_1 e_2}
{({\bf p} - {\bf p}^{\prime})^2 + R_s^2} = 
\bar{V}^C({\bf p} - {\bf p}^{\prime)}\;, 
\end{equation} 
and is taken outside the integral signs in Eqs. (31), 
\begin{equation} 
\begin{array}{rcl} 
\bar{U}({\bf p},{\bf p}^{\prime}; \epsilon) & = & \bar{V}^C({\bf p} - 
{\bf p}^{\prime}) I({\bf p},{\bf p}^{\prime}; \epsilon) + \cdots\;, 
\nonumber \\ [2mm]
\bar{W}({\bf p},{\bf p}^{\prime}; \epsilon) & = & \bar{V}^C({\bf p} - 
{\bf p}^{\prime}) J({\bf p},{\bf p}^{\prime}; \epsilon) + \cdots\;, 
\label{line2} \\ [2mm]
\bar{Z}({\bf p},{\bf p}^{\prime}; \epsilon) & = & \bar{V}^C({\bf p} - 
{\bf p}^{\prime}) L({\bf p},{\bf p}^{\prime}; \epsilon) + \cdots\;, 
\nonumber 
\end{array}
\end{equation}
where
\begin{eqnarray}
I({\bf p},{\bf p}^{\prime}; \epsilon) & = & \int \frac{d{\bf k}}
{(2\pi)^3}\frac{u({\bf k}) u\left({\bf k}-2{\bf Q}\right)}{\left[
\frac{k^2}{2\mu_{23}} + b - \epsilon_p\right] 
\left[\frac{\left({\bf k}-2{\bf Q}\right)^2}{2\mu_{23}} + b - 
\epsilon_{p'}\right]}\;, \nonumber \\ [5mm] 
J({\bf p},{\bf p}^{\prime}; \epsilon) & = & \int \frac{d{\bf k}}
{(2\pi)^3}\frac{u({\bf k}) u\left({\bf k}-2{\bf Q}\right)}{\left[
\frac{k^2}{2\mu_{23}} + b \right] 
\left[\frac{\left({\bf k}-2{\bf Q}\right)^2}{2\mu_{23}} + b - 
\epsilon_{p'}\right]}\;, \label{line2} \\ [5mm] 
L({\bf p},{\bf p}^{\prime}; \epsilon) & = & - \int \frac{d{\bf k}}
{(2\pi)^3}\frac{u({\bf k}) u\left({\bf k}-2{\bf Q}\right)}
{\left[\frac{k^2}{2\mu_{23}}+ b \right]
\left[\frac{k^2}{2\mu_{23}} + b - \epsilon_p\right] 
\left[\frac{\left({\bf k}-2{\bf Q}\right)^2}{2\mu_{23}} + b - 
\epsilon_{p'}\right]}\;. \nonumber
\end{eqnarray} 
Note, that the factors $I$, $J$ and $L$ in Eqs. (34) do not contain the
screening parameter $R_s$, they are essentially non-local and fall rapidly
at large values of the momenta  $p$ and $p^{\prime}$.  In accordance with
Eq. (27) there exists a relation between the factors (34) having the form 
\begin{equation} 
L({\bf p},{\bf p}^{\prime}; \epsilon) =\left(\epsilon-\frac{p^2}
{2\mu_1}\right)^{-1} \left[ J({\bf p},{\bf p}^{\prime}; \epsilon) - 
I({\bf p},{\bf p}^{\prime}; \epsilon)\right].  
\end{equation} 
The diagonal elements $I$, $J$ and $L$  with the momenta on the energy shell
are equal to 
\begin{eqnarray} 
I({\bf p}_0,{\bf p}_0; \epsilon) &=&  J({\bf p}_0,{\bf p}_0; \epsilon) 
 = 1\;, \nonumber \\ [5mm]
L({\bf p}_0,{\bf p}_0; \epsilon) &=& - \int \frac{d{\bf k}}{(2\pi)^3}
\frac{u^2(k)}{\left(b + \frac{k^2}{2\mu_{23}}
\right)^3}\;. \label{line2}  
\end{eqnarray}

The main singularity of the kernel of the integral equation (19) for 
the scattering wave function, which appears after removing the screening 
of the Coulomb interaction between charged particles, is caused by
superposing the singularity of the Coulomb potential acting between the
incident particle and the centre of mass of the two-body complex, 
\begin{equation}
V^C\left({\bf p} - {\bf p}^{\prime}\right) =
\frac{4\pi e_1 e_2}{\left({\bf p} - {\bf p}^{\prime}\right)^2} \; ,
\end{equation}
that (according to Eqs. (33)) is contained in the kernel $U$, on the 
singularity of the free Green function
\begin{equation}
\langle {\bf p} \mid g_0^1(\epsilon) \mid {\bf p}^{\prime} \rangle = 
(2\pi)^3 \delta\left({\bf p} - {\bf p}^{\prime}\right)\left(\epsilon - 
\frac{p^2}{2\mu_1} + i0 \right)^{-1}\;.
\end{equation}

The limiting passage to the non-screened Coulomb interaction in the
equation (19) we perform after prior reformulation of the equation by
extracting explicitly the term $g_0^1(\epsilon)\bar{V}^C$, which
corresponds to the pure Coulomb interaction, from the kernel (19). 
Transposing this term to the left side  of the equation (19) and acting 
by the inverted operator  $1 - g_0^1(\epsilon)\bar{V}^C$ on both of the
sides of the obtained equation, we write the equation for the function
$\bar{\chi}_{{\bf p}_0}$ in the following form
\begin{equation}
\bar{\chi}_{{\bf p}_0} =\bar{\psi}_{{\bf p}_0} + 
\bar{g}^C(\epsilon) [\bar{U}(\epsilon)S\left(\epsilon - b - h_0^1
\right) - \bar{V}^C]\bar{\chi}_{{\bf p}_0}\;. 
\end{equation}
The introduced scattering function $\bar{\psi}_{{\bf p}_0}^C$ for the 
screened Coulomb potential $\bar{V}^C$ (32) and the corresponding Green
function $\bar{g}^C(\epsilon)$ are determined by the equations:
\begin{eqnarray}
\bar{\psi}_{{\bf p}_0}^C & = & \varphi_{{\bf p}_0} + 
g_0^1(\epsilon) \bar{V}^C \bar{\psi}_{{\bf p}_0}^C, \nonumber \\ [2mm]   
\bar{g}^C(\epsilon) & = & g_0^1(\epsilon) + 
g_0^1(\epsilon) \bar{V}^C \bar{g}^C(\epsilon). \label{line2}
\end{eqnarray}

The passage to the non-screened Coulomb interaction is performed in the
regularized equation (39) automatically taking into account accordingly
to Refs. [5,6] that the functions $\bar{\psi}_{{\bf p}_0}^C$ and 
$\bar{\chi}_{{\bf p}_0}$ for $R_s \rightarrow \infty$ have a form of the
products of the known singular phase factor and the respective functions
$\psi_{{\bf p}_0}^C$ and $\chi_{{\bf p}_0}$, which present a finite limit
and describe the physical scattering process with the non-screened Coulomb
interaction,
\begin{eqnarray}
\bar{\psi}_{{\bf p}_0}^C({\bf p})
&\stackrel{R_s\rightarrow\infty}{\longrightarrow}&
\exp[-i\eta(\ln 2p_0R_s -\mbox{\^{C}})] 
\psi_{{\bf p}_0}^C({\bf p})\;, \nonumber \\
\bar{\chi}_{{\bf p}_0}^C({\bf p})
&\stackrel{R_s\rightarrow\infty}{\longrightarrow}&
\exp[-i\eta(\ln 2p_0R_s -\mbox{\^{C}})] 
\chi_{{\bf p}_0}^C({\bf p})\;, 
\end{eqnarray}
where
\begin{equation}
\eta = \frac{\mu_1e_1e_2}{\hbar^2p_0} = \frac{e_1e_2}{\hbar} 
\sqrt{\frac{\mu_1}{2\epsilon}} \; 
\end{equation} 
is Sommerfeld's parameter, \^{C}$ = 0.577215\ldots$ being Euler's
constant [26].  

The final integral equation for the function $\chi_{{\bf p}_0}^C({\bf p})$ 
we obtain in the form
\begin{eqnarray}
\chi_{{\bf p}_0}({\bf p}) & = & \psi_{{\bf p}_0}^C({\bf p}) 
+\int \frac{d{\bf p}^{\prime}}{(2\pi)^3} \int 
\frac{d{\bf p}^{\prime\prime}}{(2\pi)^3} \langle {\bf p} \mid 
g^C(\epsilon) \mid {\bf p}^{\prime\prime} \rangle \\ \nonumber 
&    & \cdot{}\left[U({\bf p}^{\prime\prime}, {\bf p}^{\prime}; 
\epsilon) S\left(\epsilon - b - \frac{p\prime^2}{2\mu_1}\right) - 
V^C\left({\bf p}^{\prime\prime} - {\bf p}^{\prime}\right)\right] 
\chi_{{\bf p}_0}({\bf p}^{\prime}).  \label{line2} 
\end{eqnarray}
where $ \psi_{{\bf p}_0}^C({\bf p})$ is the Coulomb scattering wave function
in the momentum space, 
\begin{equation}
\psi_{{\bf p}_0}^C({\bf p}) = C(\eta) \varphi_{{\bf p}_0}^C({\bf p})\;,
\quad \varphi_{{\bf p}_0}^C({\bf p}) = \int d\mbox{\boldmath ${\rho}$} 
e^{-i{\bf p}\mbox{\boldmath ${\rho}$}} \varphi_{{\bf p}_0^C}
(\mbox{\boldmath $\rho$}).  
\end{equation} 
In the configuration space the Coulomb scattering function
$\psi_{{\bf p}_0}^C(\mbox{\boldmath $\rho$})$ has the form [24] 
\begin{equation}
\psi_{{\bf p}_0}^C(\mbox{\boldmath $\rho$}) = 
C(\eta) \varphi_{{\bf p}_0}^C(\mbox{\boldmath $\rho$})\;,\quad  
\varphi_{{\bf p}_0}^C(\mbox{\boldmath $\rho$}) =   
e^{i{\bf p}_0\mbox{\boldmath ${\rho}$}}\left._1F_1\left(-i\eta, 1,
i\left[p_0\rho-{\bf p}_0\mbox{\boldmath $\rho$}\right]\right)\right.\;,
\end{equation} 
where $\mbox{\boldmath $\rho$} \equiv \mbox{\boldmath $\rho$}_1 =
(m_2 {\bf r}_2 + m_3{\bf r}_3)/m_{23} - {\bf r}_1$ is the relative Jacobi 
radius-vector between the centre of mass for the particles 2 and 3
and the particle 1 (${\bf r}_i$ being the radius-vector of the particle $i$), 
that corresponds to the relative momentum ${\bf p}_1$ (7). The coefficient 
\begin{equation} 
C(\eta) = exp(- \frac{1}{2} \pi \eta) \Gamma(1+ i \eta) 
\end{equation} 
in Eqs. (44) and (45) corresponds to the normalization of the incident wave on
the unit probability density
\begin{equation}
\mid \psi_{{\bf p}_0}^C(\rho\rightarrow \infty) \mid \rightarrow 1 ,
\end{equation} 
providing the known result for the probability that the particle is at the 
point $\rho=0$ (relatively to the probability that it is in the incident beam) 
in the case of the pure Coulomb scattering of two charges --- the penetration
factor for the Coulomb field of a point particle
\begin{equation} 
P_0 \;\; = \;\; \frac{ \mid \psi_{{\bf p}_0}^C(\rho=0) \mid^2} 
{\mid \psi_{{\bf p}_0}^C(\rho\rightarrow \infty) \mid^2}\;\; = \;\; 
\mid C(\eta) \mid^2 \;\; = \;\; \frac{2\pi\eta}{e^{2\pi\eta}-1}\;.  
\end{equation} 

It is obvious that the pole and Coulomb singularities in the kernel of the 
equation (43), unlike Eq. (19), do not appear in one point and hence the 
integral equation (43) is polar.

The solution of the equation (43), which expresses $\chi_{{\bf p}_0}$ 
in terms $\psi_{{\bf p}_0}^C$, can be written in the following form:     
\begin{equation}
\chi_{{\bf p}_0} = \psi_{{\bf p}_0}^C + R(\epsilon) \psi_{{\bf p}_0}^C\;,    
\end{equation}
where $R(\epsilon)$ is the resolvent of the non-screened kernel $K(\epsilon)$,
\begin{equation}
R(\epsilon) = [1 - K(\epsilon)]^{-1} - 1\;,
\end{equation}
\begin{equation}
K(\epsilon) = g^C(\epsilon) [U(\epsilon) S\left(\epsilon - b - h_0^1
\right) -V^C]\;. 
\end{equation}

The same behaviour as (41) when the screening distance tends to infinity
$(R_s \rightarrow \infty)$ is also the case for the effective function     
$\bar{\psi}_{{\bf p}_0}^{eff}$ (24),
\begin{equation}
\bar{\psi}_{{\bf p}_0}^{eff}({\bf p})
\stackrel{R_s\rightarrow\infty}{\longrightarrow}
\exp\left[-i\eta(\ln 2p_0R_s -\mbox{\^{C}})\right] 
\psi_{{\bf p}_0}^{eff}({\bf p})\;, 
\end{equation}
that follows immediately from the representation (25) of the function
$\bar{\psi}_{{\bf p}_0}^{eff}$ in terms of the function 
$\bar{\chi}_{{\bf p}_0}^{eff}$ and the behaviour (41) of the function
$\bar{\chi}_{{\bf p}_0}^{eff}$. Extracting from the expression (25) in 
the limit case $R_s\rightarrow \infty$ the singular phase factors, 
according to Eqs. (41) and (52), we obtain the expression for the effective
scattering wave function $\psi_{{\bf p}_0}^{eff}$ , that describes the
scattering process in the case of the non-screened potential
\begin{equation} 
\psi_{{\bf p}_0}^{eff}({\bf p}) = \chi_{{\bf p}_0}(\vec{p}) + 
\int \frac{d{\bf p}^{\prime}}{(2\pi)^3}  
Z({\bf p}, {\bf p}^{\prime}; \epsilon) S\left(\epsilon - b - 
\frac{p\prime^2}{2\mu_1}\right) \chi_{{\bf p}_0}({\bf p}^{\prime})\;.  
\end{equation}

The formula (53) for the effective wave function 
can be written in the form of the expression through the Coulomb wave
function $\psi_{{\bf p}_0}^C$ using the formula (49) for the function
$\chi_{{\bf p}_0}$ (and the formula (27) in the case of the non-screened 
kernel $Z$), 
\begin{equation} 
\psi_{{\bf p}_0}^{eff} = \psi_{{\bf p}_0}^C + 
g_0^1(\epsilon) \left\{W(\epsilon) S(\epsilon - b - h_0^1) [1 + 
R(\epsilon)] - V^C\right\} \psi_{{\bf p}_0}^C\;.  
\end{equation}
\\ [1mm]

\noindent {\bf 3. The probability of the penetration of a charged 
particle through the Coulomb field of a two-particle complex} \\ 

In the context of the elaborated three-particle formalism, it is of interest 
to investigate the approach of a charged particle and a charged two-particle 
bound complex, specifically, the phenomenon of the penetration of a charged 
particle through the complicated Coulomb barrier of a two-particle complex.

In the configuration space, the effective wave function can be found using 
the  Fourier transform
\begin{equation}
\psi_{{\bf p}_0}^{eff}(\mbox{\boldmath $\rho$}) = 
\int \frac{d{\bf p}}{(2\pi)^3}  
¥^{i{\bf p}\mbox{\boldmath ${\rho}$}} \psi_{{\bf p}_0}^{eff}({\bf p})\;.  
\end{equation}
The value of the effective scattering wave function at the point of the 
configuration space $\rho = 0$ can be determined according to Eq. (55) as
an integral of the Fourier image over the whole momentum space,
\begin{equation}
\psi_{{\bf p}_0}^{eff}(\rho = 0) = \int \frac{d{\bf p}}{(2\pi)^3}  
\psi_{{\bf p}_0}^{eff}({\bf p})\;.  
\end{equation}

Using the expression (54) for $\psi_{{\bf p}_0}^{eff}({\bf p})$, the formula
(56) is written in the form 
\begin{eqnarray} 
\psi_{{\bf p}_0}^{eff}(\rho = 0) & = & \psi_{{\bf p}_0}^{C}
(\rho = 0) \nonumber \\ [4mm] 
 &+&  \int \frac{d{\bf p}d{\bf p}^{\prime}}{(2\pi)^6}  
\frac{1}{\epsilon - \frac{p^2}{2\mu_1} + i0}\left\{ \int \frac{d{\bf 
p}^{\prime\prime}} {(2\pi)^3} W({\bf p}, {\bf p}^{\prime\prime};\epsilon) 
S\left(-b+\epsilon_{p''}\right)\right. \nonumber \\ [3mm] 
&\cdot& \left.\langle{\bf p}^{\prime\prime} \mid \left[1 + 
R(\epsilon)\right] \mid {\bf p}^{\prime} \rangle - 
V^C({\bf p}- {\bf p}^{\prime}) 
\rule{0mm}{6.5mm}\right\} \psi_{{\bf p}_0}^C({\bf p}^{\prime})\;.  
\label{line3} 
\end{eqnarray} 

The probability of finding a charge particle at the point of the centre of
mass of a two-particle complex (relatively to the probability of finding it  
in the incident beam) is equal to 
\begin{equation}
P = \frac{\mid\psi_{{\bf p}_0}^{eff}(\rho = 0)\mid^2}
{\mid\psi_{{\bf p}_0}^{eff}(\rho \rightarrow \infty)\mid^2}\;.   
\end{equation}
The quantity $P$ is referred to as the penetration factor for the Coulomb 
field of the two-particle complex.

In accordance with Eqs. (44) and (45), extracting from the expression 
(57) the common coefficient --- the normalized factor of the Coulomb 
scattering function $C(\eta)$ --- the formula (57) for 
$\psi_{{\bf p}_0}^{eff}(\rho = 0)$ can be given in the form 
\begin{equation} 
\psi_{{\bf p}_0}^{eff}(\rho = 0) = C(\eta) \left[1 + 
D\left(\eta, p_0\right)\right]\;, 
\end{equation} 
where the dimensionless quantity 
\begin{eqnarray} 
D\left(\eta, p_0\right) & = & \int \frac{d{\bf p}d{\bf p}^{\prime}} 
{(2\pi)^6} \frac{1}{\epsilon_p}\left\{ \int \frac{d{\bf p}^{\prime\prime}} 
{(2\pi)^3} W\left({\bf p}, {\bf p}^{\prime\prime};\epsilon\right)S\left(-b+ 
\epsilon_{p''}\right)\right.\nonumber   \\ [3mm]
& \cdot & \left.\langle{\bf p}^{\prime\prime} \mid \left[1 + 
R(\epsilon)\right] \mid {\bf p}^{\prime} \rangle - V^C\left({\bf p}- 
{\bf p}^{\prime}\right)\rule{0mm}{6.5mm}\right\}\varphi_{{\bf p}_0}^C({\bf 
p}^{\prime})\;, \label{line2} 
\end{eqnarray} 
that describes the structure effect of the complex depends on the energy 
of the relative motion of the particle and the complex
$\epsilon $, the parameter $\eta $ and two parameters which characterize
the potential of the interaction between the constituents of the complex
or the bound state of the complex, in capacity of those it is convenient
to use the binding energy $b$ and a quantity connected with the range of
the interaction.

Since the kernel of the reduced operator in parentheses of the expression
(54) does not contain the operator of the long-range Coulomb interaction,
the normalization of the effective scattering wave function
$\psi_{{\bf p}_0}^{eff}$  at infinitely large distances does not differ
from the corresponding normalization of the scattering Coulomb function (47),
\begin{equation} 
\mid\psi_{{\bf p}_0}^{eff}(\rho \rightarrow \infty)\mid \rightarrow 1\;.  
\end{equation}
Substituting the formulae (59) and (61) into Eq. (58), the expression for the
probability of the penetration of the incident charged particle into the
centre of mass of the two-particle complex (or the penetration factor for
the particle in the Coulomb field of the complex) is expressed in the form
\begin{equation} 
P = P_0 \{ 1 + 2 \mbox{Re}D + \mid D\mid^2 \}\;, 
\end{equation}
where $P_0$ is the probability of finding at one point of the incident 
particle with the charge $e_1$ and the mass $m_1$ and a point "particle" with 
the charge and mass equal to the summary charge and mass of the 
particles of the complex, $e_2$ and $m_{23}$, respectively, and located at the 
centre of mass of the complex --- the penetration factor for the charge 
particle in the Coulomb field of the structureless complex (see the formula 
(48)). The sum of the second and third terms in the parentheses of the 
expression (62) describes the relative deviation of the penetration factor for 
the charged particle in the Coulomb field of the two-particle complex from that 
of the corresponding point charge, 
\begin{equation} 
\Delta \equiv \frac{P - P_0}{P_0} = 2 \mbox{Re}D + \mid D\mid^2 \;.  
\end{equation}

In a general way, characterizing an influence of the structure of the
complex on the penetration of the point particle 1 through the Coulomb
field of the complex, the quantity $\Delta$ (63) depends on the charges and 
masses of the particles of the system, the incident momentum $p_0$ and
the parameters that describe the bound state of the two-body complex. \\ 

\noindent {\it  3.1 The relative deviation $\Delta$ in the linear
approximation in the parameter $\eta$} \\ 

Let us determine the probability of approaching a charge point particle and 
two-particle complex ussing the formula (60) in the assumption that the
parameter $\mid\eta\mid$ is small,  
\begin{equation} 
\mid\eta\mid < 1 \;.  
\end{equation}

In the linear approximation in $\eta$, taking in the formula (60) in 
accordance with Eqs. (44) and (45) (along with Eqs. (33) and (34)) only the 
first terms of expansions in the parameter $\eta$ of the function 
$\varphi_{\vec{p}_0}^C$ and the kernels $W$ and $R$ into account, we have 
\begin{equation} 
\begin{array}{c} \varphi_{{\bf p}_0}^C\left({\bf p}^{\prime}\right)= 
\varphi_{{\bf p}_0}\left({\bf p}^{\prime}\right) + O(\eta), \nonumber \\ [3mm] 
W\left({\bf p},{\bf p}_0; \epsilon\right) =  V^C\left({\bf p} - 
{\bf p}_0\right) J\left({\bf p},{\bf p}_0; \epsilon\right) + o(\eta), 
\quad J\left({\bf p},{\bf p}_0;\epsilon\right)=F_{00}\left(\mid{\bf p}- 
{\bf p}_0\mid\right) , \label{line2} \\ [3mm] R\left({\bf p}^{\prime\prime}, 
{\bf p}^{\prime};\epsilon\right)= (2\pi)^3 \delta\left({\bf p}^{\prime\prime} 
- {\bf p}^{\prime}\right) + O(\eta), \nonumber 
\end{array} 
\end{equation} 
where $F_{00}(q)$ is the formfactor of the charge distribution of the
two-particle complex (that is in a $S$-wave bound state), 
\begin{equation} 
F_{00}(q) = \int \frac{d{\bf k}}{(2\pi)^3} 
\psi_0^*\left({\bf k}\right) \psi_0\left({\bf k}-\frac{m_3}{m_{23}}{\bf 
q}\right) = \int d{\bf r} \exp\left(i\frac{m_3}{m_{23}}{\bf q}{\bf r}\right) 
\mid \psi_0\left({\bf r}\right)\mid^2, 
\end{equation} 
and the expression for the quantity $D$ is given by 
\begin{eqnarray} 
D\left(\eta, p_0\right) & = & 8\pi\eta p_0 \int \frac{d{\bf p}}{(2\pi)^3} 
\frac{1 - F_{00}\left(\mid {\bf p}-{\bf p}_0\mid\right)}{\left(p^2-p_0^2
- i0\right)\left({\bf p}-{\bf p}_0\right)^2} + o(\eta)\;, \nonumber \\ [5mm] 
& = & \eta \left[A(p_0) + iB(p_0)\right]  + o(\eta)\;, \quad\quad\quad 
\mid\eta\mid < 1 \;. \label{line2} 
\end{eqnarray} 
Here the following notations are used:
\begin{eqnarray} 
A(p_0) & = & \frac{1}{\pi} \int_{0}^{\infty} dq \frac{1 - F_{00}(q)}{q}
\ln\left(\frac{q+2p_0}{\mid q-2p_0 \mid}\right), \nonumber \\
B(p_0) & = & \int_{0}^{2p_0} dq \frac{1-F_{00}(q)}{q} , \label{line2}
\end{eqnarray}
or in terms of Eq. (66),
\begin{eqnarray} 
A(p_0) & = & \frac{\pi}{2} - 4\int_{0}^{\infty} dr r^2 |\psi_0(r)|^2
\int_{0}^{\infty}\frac{dq}{q} j_0\left(\frac{m_3}{m_{23}}qr\right)
\ln\left(\frac{q+2p_0}{\mid q-2p_0 \mid}\right), \nonumber \\
B(p_0) & = & 4\pi\int_{0}^{\infty} dr r^2 |\psi_0(r)|^2 \left[ j_(x_0) + 
\mbox{Cin}(x_0) - 1 \right]\; , \;\; 
x_0 = 2\frac{m_3}{m_{23}}p_0r\;, \label{line2}
\end{eqnarray}
where $j_0(x)$ is the spherical Bessel function, $\mbox{Cin}(x)$ is the
integral function, $\mbox{Cin}(x) = -\mbox{Ci}(x) + \ln(x) + \mbox{\^{C}}\;,
\mbox{Ci}(x)$ being the integral cosine, and \^{C} is the Euler 
constant [26].

In the limiting case of a negligible mass of the neutral constituent 
particle of the complex ($m_3 \ll m_2$), the centre of mass and the centre
of charge are coincident ($m_3/m_{23}\rightarrow 0$), the form factor
approaches unity, $F_{00}(q) \rightarrow 1$, and the quantities $A(p_0)$ 
and $B(p_0)$ (as $D$) tend to zero.

Using Eq. (67), for the ratio of the probability of penetration of a charged 
particle to the centre of mass $P$ (62) to the probability of approaching
a charged particle and a point charge (that equals to the charge of the 
complex) $P_0$ (48) we obtain 
\begin{equation} 
\frac{P}{P_0} = 1 + \Delta\; , 
\end{equation}
where the quantity
\begin{equation} 
\Delta = 2A(p_0)\eta + o(\eta)\;.  
\end{equation}
describes the deviation of the probability $P$ from $P_0$.

For the separable potential (9) with the factor $u(k)$ of the Yukawa form,
\begin{equation} 
u(k) = \frac{\sqrt{2\pi\kappa\beta(\kappa+\beta)^3}}{\mu_{23}} 
\frac{1}{k^2+\beta^2}\;,
\end{equation}
where the parameter $\beta$ characterizes the inverse of the range of 
interaction and the normalization corresponds to the normalization of
the wave function of the complex to unit (12), the form factor of the charge 
distribution of the two-body complex (66) can be written in the analytical 
form, 
\begin{eqnarray} 
F_{00}(q) & = & \frac{2m_{23}}{m_3}\frac{\kappa\beta (\beta+\kappa)}
{(\beta-\kappa)^2q}\left\{\arctan\left(\frac{1}{2}\frac{m_3}{m_{23}}
\frac{q}{\kappa}\right) \nonumber\right. \\ 
&\hfill & + \left.\arctan\left(\frac{1}{2}\frac{m_3}{m_{23}}
\frac{q}{\beta}\right) - 2\arctan \left(\frac{m_3}{m_{23}}
\frac{q}{\beta+\kappa}\right) \right\}\; . \label{line2} 
\end{eqnarray} 

Using Eq. (73), the expression for the quantity $A(p_0)$ (68) that is contained 
in Eq. (71), becomes 
\begin{eqnarray} 
A(p_0) & = & \frac{\pi}{2} - \frac{\kappa_0\beta_0(\beta_0+\kappa_0)} 
{\pi(\beta_0-\kappa_0)^2} \int_0^\infty \frac{dx}{x^2} 
\left[\arctan\left(\frac{x}{\kappa_0}\right) \nonumber\right. \\ &   & + 
\left.\arctan\left(\frac{x}{\beta_0}\right) - 2\arctan 
\left(\frac{2x}{\beta_0+\kappa_0}\right) \right] \ln\left(\frac{x+1}{\mid 
x-1\mid}\right)\; , \label{line2} 
\end{eqnarray} 
where 
\begin{equation} 
\kappa_0 = \frac{m_{23}}{m_3}\frac{\kappa}{p_0}\;,\quad 
\quad \beta_0 = \frac{m_{23}}{m_3}\frac{\beta}{p_0}\;.  
\end{equation} 
The quantity $A$ (68) depends on the quantities which describe the bound 
state of the two-particle complex --- on the binding energy 
$b=\kappa^2/2\mu_{23}$ and the parameter $\beta$ for the potential
with the factor (72) --- and also on the incident momentum of the relative 
motion of the particle and the centre of mass of the complex $p_0$ (or the
energy $\epsilon=p_0^2/2\mu_1$). Acording to Eq. (74), it as dimensionless 
quantity is constructed from two dimensionless parameters, $\kappa_0$ and
$\beta_0$ (75). In view of the fact that the expression (74) contains a  
linear combination of three like integrals, the quantity $A$ is written as 
\begin{equation} 
A_(p_0) = \frac{\pi}{2} - \frac{\kappa_0\beta_0 
(\beta_0+\kappa_0)} {\pi(\beta_0-\kappa_0)^2} 
\left[a\left(\frac{1}{\kappa_0}\right) + a\left(\frac{1}{\beta_0}\right) - 
2a\left(\frac{2}{\kappa_0+\beta_0}\right)\right]\;,
\end{equation}
where            
\begin{equation} 
a_(\xi) = \int_0^\infty \frac{dx}{x^2}
\arctan (\xi x ) \ln \left(\frac{x+1}{|x-1|}\right)\;.
\end{equation}

We note that in the case of the potential (9), (72) the wave function of
the complex in the bound state (12) having a physical behaviour at 
asymptotically large distances between the particles is possible only provided 
that the unequality 
\begin{equation} 
\kappa < \beta\;.  
\end{equation} 
is obeyed. It immediately follow the unequalities between the values of 
the argument  of the function $a(\xi)$, which are contained in the 
expression (76):  
\begin{equation} 
\frac{1}{\beta_0} < \frac{2}{\kappa_0+\beta_0} < \frac{1}{\kappa_0}\;.  
\end{equation}

In a more compact form, simplifying the expression for linear combination
of the arctangent functions with different arguments in Eq. (74) and taking
into account the value of the integral
\begin{equation} 
\int_0^\infty \frac{dx}{x} \ln \frac{x+1}{\mid x-1 \mid} = \frac{\pi^2}{2}\;,
\end{equation}
the formula (74) can be written as
\begin{equation} 
A(p_0) = \frac{2}{\pi} \eta \int_0^\infty \frac{dx}{x^2}
\ln \left(\frac{x+1}{|x-1|}\right) \Phi (x,p_0)\;,
\end{equation}
where
\begin{equation} 
\Phi (x,p_0) = x - \frac{d}{f} \arctan \left(\frac{fx}{4x^4+cx^2+d} 
\right)\; , \label{line1} 
\end{equation}
\begin{displaymath}
f=(\beta_0+\kappa_0)(\beta_0-\kappa_0)^2\; , \quad
c=3\beta_0^2+2\beta_0\kappa_0+3\kappa_0^2\; ,\quad
d=\beta_0\kappa_0(\beta_0+\kappa_0)^2\; . 
\end{displaymath}

From the formulae (81) and (82) it follows at once that the function $A(p_0)$ 
is positive for all positive values of the argument, since the function
$\Phi(x,p_0)$ and the integrand in Eq. (81) take positive values for all
$x>0$ (at the point $x=0$ the function $\Phi$ and its first and second 
derivatives $\Phi'$ and $\Phi"$ are vanished).

As a consequence, in the first approximation of the expansion of $P/P_0$ in
$\eta$ (70), the correction for the structure of the complex $\Delta$ (71)
is positive ($\Delta > 0$) in the case of the repulsive Coulomb interaction
between charged particles ($e_1e_2 > 0$) and negative  ($\Delta < 0$) in the 
case of the attractive Coulomb interaction ($e_1e_2 < 0$). Hence, allowance for 
the structure of complex results in the penetration factor to increase for like 
charges of the particle and the complex whereas to decrease for unlike charges,
\begin{equation} 
P > P_0\;, \;\; \mbox{ ïªé® }\;\; e_1e_2 > 0\;,\;\; \mbox{ ÷ }\;\; 
P < P_0\;, \;\; \mbox{ ïªé® } \;\; e_1e_2 < 0\;.
\end{equation}
In this approximation, the correction for the structure of the complex 
$\Delta$ is proportional to the reduced mass of the incident particle and
the particles of the complex $\mu_1$ (the inverse Bohr radius of an atom
composed of the incident particle and the two-particle complex). \\ [1.2cm]

\noindent {\it  3.2 A simple analytical estimation of $\Delta$ in a restricted
range of small values of $p_0$} \\ 

The magnitude of the influence of the structure of the two-fragment complex
on the probability of the penetration of a charged particle through the
Coulomb field of the complex can be easily estimated in the event that the
incident momentum $p_0$  is small relative to $\kappa$ such that
\begin{equation} 
\frac{1}{\kappa_0} = \frac{m_3}{m_{23}} \frac{p_0}{\kappa} < 1\;.
\end{equation}
Then it becomes possible to perform an approximate integration in Eq. (76) in
an explicit form. We restrict ourselves to taking into account the first four 
terms in the expansion of the function $a(\xi)$ (76) in the power series in  
$\xi$, 
\begin{equation} 
a(\xi) = a(0) + \xi a'(0) + \frac{1}{2} \xi^2 a''(0) + o(\xi^2)\;, 
\end{equation} 
Splitting the interval of the integration in the expressions for the function
$a(\xi)$ (76) and its derivatives into two parts --- from $0$ to $1$ and
from $1$ to $\infty$ --- and performing the integration with the help of 
Eq. (80) and the formulae [27] 
\begin{eqnarray}
\int_0^1 dx x \ln \left(\frac{1+x}{1-x}\right) = 1 \;\;,\quad& &\int_0^1 dx 
\frac{x}{x^2+\xi^2} \ln \left(\frac{1+x}{1-x}\right) = 
\left[\arctan\left(\frac{1}{\xi}\right)\right]^2 , \nonumber \\
\int_0^1 dx \frac{x}{(x^2+\xi^2)^2} \ln \left(\frac{1+x}{1-x}\right)& = 
&\frac{1}{\xi (1+\xi^2)}\arctan\left(\frac{1}{\xi}\right) \;, \label{line2} \\
\int_0^1 dx \frac{x}{(x^2+\xi^2)^3} \ln \left(\frac{1+x}{1-x}\right)& = 
&\frac{1}{4\xi^3 (1+\xi^2)^3}\left[\left(1+3\xi^2\right)\arctan\left(\frac{1}
{\xi}\right)+\xi\right] \;, \nonumber \\ 
\int_0^1 dx \frac{x^3}{(x^2+\xi^2)^3} \ln \left(\frac{1+x}{1-x}\right)& = 
&\frac{1}{4\xi (1+\xi^2)^2}\left[\left(3+\xi^2\right)\arctan\left(
\frac{1}{\xi}\right)-\xi\right] \;, \nonumber 
\end{eqnarray}
we then find the coefficients of the expansion (85):  
\begin{equation} 
a(0)=0\;,\quad a'(0)=\frac{\pi^2}{2}\;,\quad a''(0)=-\pi\;,\quad a'''(0)=0\;.  
\end{equation} 
As a result, the approximated expression for the function $a(\xi)$ (85) takes the form 
\begin{equation} 
a(\xi) = \frac{\pi^2}{2} \xi - \frac{\pi}{2} \xi^2 + O(\xi^3)\;,\quad\quad 
0<\xi<1\;.  
\end{equation}
In the formula (88) the expansion of the function $a(\xi)$ (76) is performed
in the interval of convergence $ 0<\xi<1 $. It is worth noting that the 
obtained expression (88) does not contradict to the odd parity of the function 
$a(\xi)$ (76), since the second derivative of the function $a''(\xi)$ has
a finite discontinuity at the point $\xi=0$,
\[
a''(\xi) \longrightarrow -\mbox{sgn}(\xi)\pi\;, \mbox{ when } \xi 
\rightarrow 0\; ,             
\]
taking the values of opposite signs when approaching $0$ from the positive and 
negative values of the argument $\xi$, $a''(0_{\pm})=\mp\pi$. 
(Here, $\mbox{sgn}(\xi)$ is the sign function of $\xi$.)

Substituting to Eq. (76) the first terms of the expansion of the function 
$a(\xi)$ for all three values of the argument in Eq. (76), which in accordance 
to the inequalities (79) and (84) are smaller than $1$, we wright the main 
(linear with respect to $p_0$) term of the expansion of the quantity $A$ in the 
form 
\begin{eqnarray} A(p_0) & = & \frac{1}{2} \left(\frac{1}{\kappa_0} 
+ \frac{2}{\kappa_0+\beta_0} + \frac{1}{\beta_0}\right) + 
o\left(\frac{1}{\kappa_0}\right)\;, \nonumber \\ [3mm] & = & \frac{1}{2} 
\frac{m_3}{m_{23}} \frac{p_0}{\kappa} f\left(\beta,\kappa\right) + 
o\left(\frac{1}{\kappa_0}\right)\;, \label{line2} 
\end{eqnarray} 
where 
\begin{equation} 
f(\beta, \kappa) = \frac{\beta^2+4\beta\kappa+\kappa^2}{\beta(\beta+\kappa)}\;,
\end{equation}
Introducing the mean radius of the two-particle complex in the considered
case of the interaction (9),(72),
\begin{equation} 
R \; = \;\frac{1}{2}\langle\psi_0 \mid r \mid \psi_0\rangle \; =
\;\;\frac{1}{4\kappa} f\left(\kappa, \beta\right)\;, 
\end{equation}
we rewrite Eq. (89) as								
\begin{equation} 
A(p_0) = \frac{2m_3}{m_{23}}R p_0 + o\left(\frac{1}{\kappa_0^2}\right)\;.  
\end{equation} 
Notice that the expression (89) immediately follows also from the formula for
$A(p_0)$ (69), if in it to apply the expansion of the spherical Bessel 
function in small values of the argument, that is equivalent to the expansion
(88) of the function $a(\xi)$ in Eq. (76) in small $\xi$.

In the employed approximation, the relative deviation of the penetration factor
for the Coulomb field of the two-particle complex $P$ from the penetration
factor for the Coulomb field of the corresponding point charge $P_0$ (74) 
does not depend on the incident momentum $p_0$,
\begin{equation} 
\Delta = \frac{m_1m_3}{M} \frac{e_1e_2}{\hbar^2 \kappa} 
f\left(\kappa,\beta\right)\; = 4 \frac{m_1m_3}{M} \frac{e_1e_2}{\hbar^2}R\;, 
\end{equation}
turning out to be proportional to the ratio of the mean radius of the 
two-particle complex to Bohr's radius of an atom composed of the incident 
particle and the complex, $a_B \equiv \hbar^2/\mu_1\vert e_1e_2\vert$ , 
\begin{equation} 
\Delta = \mbox{sgn}(e_1e_2)4\frac{m_3}{m_{23}}\frac{R}{a_B}\;. 
\end{equation}

In the limiting case of the zero-range interaction between the constituent
particles 2 and 3 of bound complex that follows from the model of the
separable potential (9), (72) when the parameter $\beta$ becomes infinitely
large ($\beta \rightarrow \infty$), the factor (90) tends to $1$,
$f(\kappa,\beta)\rightarrow 1$, $R \rightarrow 1/4\kappa$, and the correction 
for structure $\Delta$ takes the form  
\begin{equation} 
\Delta \rightarrow \Delta_{zr} \;=  
\; \frac{m_1m_3}{M} \frac{e_1e_2}{\hbar^2 \kappa}\;, 
\end{equation} 

The interval of applicability of the formula (93) is limited by the values
of the relative incident momentum $p_0$ that satisfy simultaneously to the 
conditions (64) and (84), providing the convergence of the expansions in series
of the function $\Delta(\eta)$ in the parameter $\eta$ (42) and the function
$a(\frac{1}{\kappa_0})$ in the parameter $\frac{1}{\kappa_0}$ (75), and is 
determined by the inequalities
\begin{equation} 
\frac{1}{a_B}< p_0 < \frac{m_{23}}{m_3} \kappa\;.
\end{equation}
The associated values of the energy $\epsilon=p_0^2/2\mu_1$ are limited by
the inequalities
\begin{equation} 
\epsilon_R< \epsilon < \frac{m_2M}{m_1m_3} b\;, 
\end{equation} 
where the lower boundary is Rydberg's energy for an atom of the charged
incident particle 1 and the two-particle complex, $\epsilon_R=\mu_1
\mid e_1e_2\mid^2/2 \hbar^2$, and the upper bound is proportional to the 
binding energy of the complex $b$. \\ [1mm]

\noindent {\bf 4. Numerical results and discussion} \\ 

Applying the above-described formalism by the use of the formulae (71), 
(81) and (82)), we have performed calculations of the deviation of the 
relative penetration factor, $\Delta$, in specific cases of collision 
between a charged incident particle of various mass (the muon, the pion, 
the kaon and the proton) and the deuteron and between the proton and 
the lightest hypernuclei, $^3_{\Lambda}\mbox{H}$ and $^5_{\Lambda}\mbox{He}$, 
which have a well-marked two-fragment structure.

To describe the deuteron (as a system of the proton and the neutron in
the bound $S$-state), the following values of the parameters of the
potential (9),(72) $\kappa\equiv\kappa_{pn}$ and  $\beta\equiv\beta_{pn}$ 
that correspond to the experimental values of the deuteron binding 
energy $b\equiv b_{pn}= 2.224575(9)$ δV [28] and the triplet neutron-proton
scattering length $a_{np}^t= 5.424(3)$ fm [29] were used  
\begin{equation} 
\kappa_{pn}=0.23161\;\;\mbox{fm}^{-1}\;,\; 
\beta_{pn}=1.3906\;\;\mbox{fm}^{-1}\;.  
\end{equation}

In a similar way we fit the parameters of the potentials of the form (9),
(72) that support the hypernuclei $^3_{\Lambda}\mbox{H}$ and 
$^5_{\Lambda}\mbox{He}$. The parameters $\kappa\equiv\kappa_{d\Lambda}$ and
$\beta\equiv\beta_{d\Lambda}$ providing the existence of the bound S-state 
of the hypertriton $^3_{\Lambda}\mbox{H}$ as a two-fragment system that
consists of the deuteron and the $\Lambda$-hyperon, we find based on the
experimental value of the binding energy of the hypertriton,
$b\equiv b_{d\Lambda}= 0.13(5)$ MeV [30], and the value of the doublet
$\Lambda$-hyperon-deuteron scattering length $a_{d\Lambda}^d$ = 15.9 fm.
The latter follows from the correlation dependence $a_{d\Lambda}^d$ on
$b_{d\Lambda}$ that has been calculated in the paper [31] (see also Ref. [32]) 
on the basis of the three-particle description of the hypertriton as a system 
of the proton, the neutron and the $\Lambda$-hyperon with the use of the
known experimental low-energy data on p-n and $\Lambda-$nucleon interaction. 
We find in this case
\begin{equation} 
\kappa_{d\Lambda}=0.06834\;\;\mbox{fm}^{-1}\,,\;\;  
\beta_{d\Lambda}=1.1938\;\;\mbox{fm}^{-1}\;.  
\end{equation}

For the hypernucleus $^5_{\Lambda}\mbox{He}$ as a two-fragment system
composed of the $\alpha$-particle and $\Lambda$-hyperon, we find the values
of the parameter $\kappa\equiv\kappa_{\alpha\Lambda}$ that corresponds to
the experimental binding energy of the nucleus $^5_{\Lambda}\mbox{He}$, 
$b=b_{\alpha\Lambda}=3.12(2)$ MeV [30], and the parameter  
$\beta\equiv\beta_{\alpha\Lambda},$ that (together with the experimental 
$b_{\alpha\Lambda}$) reproduces the value of the root-mean-square radius 
of the bound system $^5_{\Lambda}\mbox{He}$, $\langle r^2 
\rangle_{\alpha\Lambda}^{1/2} = 2.43$ fm, calculated with the use of a
potential model of $\alpha-\Lambda$ interaction proposed in the paper [33], 
\begin{equation} 
\kappa_{\alpha \Lambda}=0.37094\;\;\mbox{fm}^{-1}\,, \;\; 
\beta_{\alpha\Lambda}= 2.177\;\;\mbox{fm}^{-1}\;.  
\end{equation}

The results of our calculation of the function $A(p_0)$, which characterizes 
the structure of each one of the considered two-fragment nuclear systems, the
deuteron and two hypernuclei, $^3_{\Lambda}\mbox{H}$ and 
$^5_{\Lambda}\mbox{He}$, are shown in Fig.1. The solid lines in the figure are
obtained according to the formulae (81) and (82), which correspond to the
separable potential of interaction between fragments of the nuclear complex
(9), (72) with the parameters (98)-(100), while the dotted lines are obtained
for the zero-range interaction ($\beta \rightarrow \infty$). It follows
from the definition (76),(77) that the function $A(p_0)$ rises as the
incident momentum $p_0$ increases. The value of $A(p_0)$ depends essentially 
on the binding energy of the two-fragment system, increasing when going to
a system with a lesser binding energy. Hence, the curve $A(p_0)$ for the
deuteron is shifted up relative to the curve for the nucleus  
$^5_{\Lambda}\mbox{He}$, as does the curve for the nucleus
$^3_{\Lambda}\mbox{H}$ relative to the curve for the deuteron. From the
side of small values  of $p_0$ in the range (96), the behaviour of $A(p_0)$  
appears to be near linear in accordance with the result of our approximate
estimate (92), the slope of the straght line being proportional to the
radius of the corresponding two-fragment complex $R$. For each nuclear 
complex the curve $A(p_0)$ calculated with the finite-range interaction is 
shifted up in comparison of that calculated with the zero-range interaction.

In the Table 1 and the Figures 2-4, we present our numerical results for the 
relative deviation of the penetration factor for the Coulomb field of the 
two-fragment nucleus from the penetration factor for the Coulomb field of
the corresponding point charge, $\Delta$, in relation to the incident
momentum  $p_0$, which have been obtained by using the expressions (71), 
(81) and (82)
\footnote[2]{Note that the data for $\Delta$ reported in the 
table of Ref. [34] are erroneous (except the results in the last line of the 
table). The corresponding corrected results are given in Table 1 of this 
paper.}.  
The calculations have been performed for the charged incident particle 
of various mass (the muon, the pion, the kaon and the proton) colliding with 
the deuteron and for the proton colliding with the hypernuclei 
$^3_{\Lambda}\mbox{H}$ and $^5_{\Lambda}\mbox{He}$.

The obtained values of $\Delta$ for $\mu-\mbox{d}$, $\pi-\mbox{d}$, 
$\mbox{K}-\mbox{d}$ and $\mbox{p}-\mbox{d}$ interactions differ from each 
other, since, to begin with, the incident particles have different masses, 
$m_1$, in this case $\Delta$ increases with increasing $\mu_1$. Also, as the 
data of the Table 1 and the Fig. 4 indicate, the quantity $\Delta$ depends 
essentially on the binding energy of the two-fragment nucleus $b$ increasing 
with decreasing $b$.  Comparing the results for $\mbox{p}-\mbox{d}$, 
$\mbox{p}-^3_{\Lambda}\mbox{H}$ and $\mbox{p}-^5_{\Lambda}\mbox{He}$ 
interactions between themselves, it should be remembered that in this case an 
additional doubling $\Delta$ for $\mbox{p}-^5_{\Lambda}\mbox{He}$ caused by 
that the charged fragment of the hypernucleus $^5_{\Lambda}\mbox{He}$ (the 
$\alpha$-particle $^4_2\mbox{He}$ ) has its charge twice as large as the 
charged fragments of the deuteron or the hypertriton.

For comparison, in the Table 2 we give the values of the relative deviation
$\Delta$ calculated in the simple approximation (93) which is independent
of the incident momentum $p_0$ and may be applied for values of the momentum 
$p_0$ in the range (96). The result for $\Delta$ in the explicit analytical
form (93) is obtained by approximate integration in the formula (76) using
the expansion of the function (77) in the powers of $\xi$ (88) with the
parameter $\xi$ taking the values
\begin{displaymath} 
\frac{m_3}{m_{23}}\frac{p_0}{\kappa},\quad 2\frac{m_3}{m_{23}}
\frac{p_0}{\kappa+\beta},\quad \mbox{ or }\frac{m_3}{m_{23}}
\frac{p_0}{\beta}\; .  
\end{displaymath} 
The greater the binding energy of the two-fragment system, the more it is 
accurate the approximate integration and the wider it is the range of values
of the momentum $p_0$ for which this approximation is valid. This is evident 
comparing the data in the Table 2 with the data of the direct numerical
integration listed in the Table 1: the mentioned approximation is appropriate 
in the widest range of $p_0$ for the most strongly bounded two-fragment nucleus 
$^5_{\Lambda}\mbox{He}$ .

Notice that the calculation data obtained in this work embrace energies 
both below and above the Coulomb barrier of the electrostatic repulsion 
energy between the positively charged projectile and nucleus 
$\epsilon^b=\mid e_1e_2\mid / R$. (The values of the momentum $p_0^b$ that 
correspond to the respective energies $\epsilon^b\equiv(p_0^b)^2/2\mu_1$ are 
given in the Table 2). \\[1mm]

\noindent {\bf 5. Summary} \\ 

This paper is devoted to formulation of a consistent three-body approach to
the description of the phenomenon of the penetration of a charged particle 
through the Coulomb field of a two-fragment (nuclear) complex consisted of one 
charged and one neutral particles in a bound state. The corresponding 
formalism has been constructed  on the basis of the use of the three-body 
Faddeev method and the Watson-Feshbach technique for introducing the 
effective (optical) potential of the interaction between a particle and 
a bound system.

In view of the presence of the long-range Coulomb interaction, the 
corresponding Faddeev integral equations are non-Fredholm above the threshold 
of the elastic particle-complex scattering, in this connection our study 
starts  with the use of the screened Coulomb interaction. The passage to
the limit of the non-screened Coulomb potential and the regularization of 
the obtained equations have been performed in accordance with the 
Gorshkov-Vesselova recipe by way of the explicit isolation and removal of
the known singular exponential Coulomb factor.

The three-body formalism that describes the penetration of the charged 
particle through the Coulomb field of a two-particle complex has been worked 
out in the general case for both attractive and repulsive fields. The
formula for the relative deviation $\Delta$ of the penetration factor
for the Coulomb field of the two-particle complex from the penetration factor 
for the Coulomb field of the corresponding point charge has been derived.

As a first step to the application of the developed formalism, we have derived 
a simple analytic expression for the relative deviation $\Delta$ by 
approximation the function $D(\eta,p_0)$ with the linear term of the expansion
of $D$ as a power series in $\eta$, the Sommerfeld parameter for the Coulomb 
interaction between the particle and the centre of mass of the complex, 
when $\mid\eta\mid<1$. In this case, the correction for the structure 
of the complex, $\Delta$, was found to be positive for the repulsive Coulomb 
interaction between the charged particle and negative for the attractive 
Coulomb interaction. By way of illustration, the influence of the nucleus 
structure on the probability of the coming together of a charged particle 
(the muon, the pion, the kaon and the proton) and a two-fragment nucleus 
(the deuteron and the $\Lambda$-hypernuclei $^3_{\Lambda}\mbox{H}$ and
$^5_{\Lambda}\mbox{He}$) has been studied. The calculated values of the 
relative deviation $\Delta$ (by the formulae (71),(81) and (82)) are given 
in the Table 1 and shown in the Figures 2--4. The quantity $\Delta$ is
proportional to the mass of the incident particle $m_1$ and to the charges
of the incident particle and complex, $e_1$ and $e_2$, it depends essentially 
on the binding energy of the complex $b$, increasing with decreasing $b$.
It has been found that the value of the correction for the structure, 
$\Delta$, calculated according to the model of the finite-range nuclear 
interaction exceeds noticeably the value of the corresponding correction 
for the structure $\Delta_{zr}$, calculated with the use of the model of the 
zero-range interaction, $\mid\Delta\mid > \mid\Delta_{zr}\mid$ (see the data 
of the Tables 1 and 2).

A simple estimation of the influence of the two-fragment complex on the 
probability of the penetration of a charged particle through the Gamow 
field of the complex has been made for a range of small values of the 
incident momentum $p_0$ (relative to the momentum $\kappa$ that corresponds 
to the binding energy of the two-body complex, $b$). By expanding the function
$a(\xi)$ in Eq. (76) in the parameter $\xi$ with the proviso that $\xi<1$ and 
taking the first four terms of the expansion into consideration, the function 
$\Delta(p_0)$ has been approximated by the expression (94). In this 
approximation, the function $\Delta(p_0)$ appears to be independent of $p_0$ 
and the magnitude of $\Delta$ is determined by the ratio of the mean radius 
of the two-particle complex $R$ to the Bohr radius of an atom composing of 
the incident particle and complex, $a_B$.

The three-body formalism given in this paper can be generalized to other 
cases when both of fragments of the complex are charged and thus applied to 
research the penetration of charged particles through the Coulomb fields 
of nuclear, atomic and molecular systems.

\pagebreak

\begin{itemize}
\item[{1}] G.A. Gamow, Z. Phys. 51 (1928) 204; 52 (1928) 510.
\item[{2}] E.U. Condon and R.W. Gurney, Phys. Rev. 33 (1929) 127.
\item[{3}] S.P. Merkuriev and L.D. Faddeev, Quantum Theory of Scattering
            for Few-Body Systems (Nauka, Moscow, 1985).
\item[{4}] L.D. Faddeev, Zh. Eksp. Teor. Fiz. 39 (1960) 1459.
\item[{5}] V.G. Gorshkov, Zh. Eksp. Teor. Fiz. 40 (1961) 1481.
\item[{6}] A.M. Vesselova, Teor. Mat. Fiz. 3 (1970) 326.
\item[{7}] V.F. Kharchenko and S.A. Storozhenko, Integral equations for 
			  three-nucleon problem with Coulomb interaction. Proton-deuteron
			  scattering, Preprint ITP-75-53E (Institute for Theoretical Physics, 
			  Kyiv, 1975).  
\item[{8}] E.O. Alt, P. Grassberger and W. Sandhas,  Nucl. Phys. B 2 (1967) 
           167.  
\item[{9}] E.O. Alt, W. Sandhas, H. Zankel and H. Ziegelmann, Phys. Rev. Lett. 
           37 (1976) 1537.  
\item[{10}] E.O. Alt, W. Sandhas, H. Ziegelmann, Phys. Rev. C 17 (1978) 1981.  
\item[{11}] E.O. Alt, {\it in} Dynamics of Few Body Systems, eds. Gy. Bencze, 
           P. Doleschall and J. R\'{e}vai (KFKI, Budapest, 1986) p. 367.  
\item[{12}] V.F. Kharchenko, M.A. Navrotsky and S.A. Shadchin, Nucl. Phys.  
           A 512 (1990) 294.  
\item[{13}] V.F. Kharchenko, M.A. Navrotsky and P.A. Katerinchuk, Nucl. Phys. 
           A 552 (1993) 378.  
\item[{14}] C.R. Chen, G.L. Payne, J.L. Friar and B.F. Gibson, Rhys. Rev. C 39 
           (1989) 1261; C 44 (1991) 50.  
\item[{15}] V.F. Kharchenko and S.A. Shadchin, Few Body Systems 6 (1989) 45.  
\item[{16}] V.F. Kharchenko and S.A. Shadchin, Three-body theory of the 
           effective interaction between a particle and a two-particle bound 
           system, Preprint ITP-93-24E (Institute for Theoretical Physics, 
           Kyiv, 1993).  
\item[{17}] V.F. Kharchenko and S.O. Shadchin, Ukrainian J. Phys. 42 (1997) 
           912.  
\item[{18}] E.O. Alt and A.M. Mukhamedzhanov, Phys. Rev. A 51 (1995) 3852.  
\item[{19}] V.F. Kharchenko, Ukrainian J. Phys. 45 (2000) 616.  
\item[{20}] V.F. Kharchenko, J. Phys. Studies 4 (2000) 245.  
\item[{21}] J.V. Noble, Phys. Rev. 161 (1967) 945.  
\item[{22}] N.C. Francis and K.M. Watson, Phys.Rev. 92 (1953) 291.  
\item[{23}] H. Feshbach, Ann. Phys. (N.Y.) 5 (1958) 357; 19 (1962) 287.  
\item[{24}] C.J. Joachain, Quantum Collision Theory (North-Holland -- 
           American Elsevier, Amsterdam -- New-York, 1975).  
\item[{25}] V.F. Kharchenko and S.O. Shadchin, Ukrainian J. Phys. 42 (1997) 11.
\item[{26}] Handbook of Mathematical Functions, eds. M.Abramowitz and I.A. 
           Stegun (National Bureau of Standards, Applied Mathematics 
           Series-55, 1964).  
\item[{27}] I.S. Gradshtein and I.M. Ryzhik, Tables of Integrals, Sums, Series 
           and Products (Gos. Izd. Fiz. Mat. Lit., Moscow, 1962). 
\item[{28}] '. van der Leun and C. Alderliesten, Nucl. Phys. A 380 (1982) 261. 
\item[{29}] L. Koester, W. Nistler, Z. Phys. A 272 (1975) 189.  
\item[{30}] M. Juri\'{c} et al., Nucl. Phys. B 52 (1973) 1.  
\item[{31}] V.V. Peresypkin and N.M. Petrov, Binding energy of hypertriton 
           and doublet scattering length of $\Lambda$-hyperon-deuteron 
			  scattering for nonlocal separable potentials (Institute for 
			  Theoretical Physics, Preprint ITF-75-39R, Kyiv, 1975).
\item[{32}] N.M. Petrov, Yad. Fiz. 48 (1988) 50.  
\item[{33}] Y. Kurihara, Y. Akaishi and H. Tanaka, Phys. Rev. C 31 (1985) 971.  
\item[{34}] V.F. Kharchenko and A.V. Kharchenko, Ukrainian J. Phys. 48 (2003) 
           775.  
  
\end{itemize}

\pagebreak

\begin{table}
\caption\noindent{\small Values of the relative deviation $\Delta$ 
calculated for the separable potential model (9) and (72) by the formulae (71), 
(81) and (82). The numbers in curly brackets are the values of the relative 
deviation for the zero-range interaction ($\beta\rightarrow\infty$), 
$\Delta_{zr}$} \\[5mm] 
\label{tabone}
\begin{tabular}{ccccccc} \hline 
\multicolumn{1}{c}{$p_0$, fm$^{-1}$} &
\multicolumn{1}{c}{$\mu^{\pm}-\mbox{d}$}& 
\multicolumn{1}{c}{$\pi^{\pm}-\mbox{d}$}&
\multicolumn{1}{c}{$\mbox{K}^{\pm}-\mbox{d}$}&
\multicolumn{1}{c}{$\mbox{p}-\mbox{d}$}& 
\multicolumn{1}{c}{$\mbox{p}-^3_\Lambda\mbox{H}$}&
\multicolumn{1}{c}{$\mbox{p}-^5_\Lambda\mbox{He}$}\\
\hline 
0.01 &$\pm$0.01160&$\pm$0.01507&\hspace{\fill}&\hspace{\fill}
&\hspace{\fill}& \hspace{\fill}\\ 
\hspace*{\fill}&$\{\pm$0.00799\}&$\{\pm$0.01038\}&\hspace{\fill}&\hspace{\fill}
&\hspace{\fill}& \hspace{\fill}\\ [2mm]
0.05 &$\pm$0.01158&$\pm$0.01504&$\pm$0.04525&\hspace{\fill}&\hspace{\fill}
&\hspace{\fill}\\ 
\hspace*{\fill}&$\{\pm$0.00798\}&$\{\pm$0.01036\}&$\{\pm$0.03117\}&
\hspace{\fill}&\hspace{\fill}& \hspace{\fill}\\ [2mm]
0.10 &$\pm$0.01150&$\pm$0.01494 &$\pm$0.04496&0.07195&0.16037&0.05272\\ 
\hspace*{\fill}&$\{\pm$0.00793\}&$\{\pm$0.01030\}&$\{\pm$0.03099\}&
\{0.04960\}&\{0.13772\}&\{0.03608\}\\ [2mm]
0.30 &$\pm$0.01081&$\pm$0.01403&$\pm$0.04223&0.06759&0.12737&0.05242\\ 
\hspace*{\fill}&$\{\pm$0.00751\}&$\{\pm$0.00975\}&$\{\pm$0.02935\}&
\{0.04697\}&\{0.10997\}&\{0.03590\}\\ [2mm]
0.50 &$\pm$0.00979&$\pm$0.01272&$\pm$0.03828&0.06126&0.10072&0.05183\\ 
\hspace*{\fill}&$\{\pm$0.00690\}&$\{\pm$0.00896\}&$\{\pm$0.02695\}&
\{0.04313\}&\{0.08754\}&\{0.03554\}\\ [2mm]
0.70 &\hspace{\fill}&\hspace{\fill}&$\pm$0.03428&0.05486&0.08240&0.05099\\ 
\hspace*{\fill}&\hspace{\fill}&\hspace{\fill}&$\{\pm$0.02450\}&\{0.03921\}&
\{0.07209\}&\{0.03504\}\\ [2mm]
1.00 &\hspace{\fill}&\hspace{\fill}&$\pm$0.02907&0.04653&0.06433&0.04938\\ 
\hspace*{\fill}&\hspace{\fill}&\hspace{\fill}&$\{\pm$0.02127\}&\{0.03405\}&
\{0.05682\}&\{0.03408\}\\ [2mm]
2.00 &\hspace{\fill}&\hspace{\fill}&$\pm$0.01846&0.02954&0.03673&0.04271\\ 
\hspace*{\fill}&\hspace{\fill}&\hspace{\fill}&$\{\pm$0.01444\}&\{0.02311\}&
\{0.03328\}&\{0.03007\}\\ [2mm]
3.00 &\hspace{\fill}&\hspace{\fill}&\hspace{\fill}&0.02125&0.02557&0.03637\\
\hspace*{\fill}&\hspace{\fill}&\hspace{\fill}&\hspace{\fill}&\{0.01739\}&
\{0.02358\}&\{0.02621\}\\ [2mm]
4.00 &\hspace{\fill}&\hspace{\fill}&\hspace{\fill}&0.01652&0.01958&0.03121\\ 
\hspace*{\fill}&\hspace{\fill}&\hspace{\fill}&\hspace{\fill}&\{0.01394\}&
\{0.01827\}&\{0.02301\}\\ [2mm]
\hline \vspace{5mm}
\end{tabular}
\end{table}


\begin{table}
\caption\noindent{\small Values of the relative deviation $\Delta$ 
(for the separable potential (9) and (72)) and $\Delta_{zr}$ 
(for the zero-range interaction ($\beta\rightarrow\infty$))
calculated by the approximate formula (93). The lower and upper
bounds of the interval (96) of the values of the incident momentum 
$p_0$ for which the formula (93) is applicable, $1/a_B$ and
$(m_{23}/m_3)\kappa$, are given in fm$^{-1}$. The momentum 
$p_0^b$ corresponds to the Coulomb barrier energy $\epsilon^b=
\mid e_1e_2\mid /R$} \\[5mm] 
\label{tabtwo}
\begin{tabular}{ccccccc} \hline 
\multicolumn{1}{c}{Particle-nucleus} &
\multicolumn{1}{c}{$\mu^{\pm}-\mbox{d}$}& 
\multicolumn{1}{c}{$\pi^{\pm}-\mbox{d}$}&
\multicolumn{1}{c}{$\mbox{K}^{\pm}-\mbox{d}$}&
\multicolumn{1}{c}{p-d}& 
\multicolumn{1}{c}{$\mbox{p}-^3_\Lambda\mbox{H}$}&
\multicolumn{1}{c}{$\mbox{p}-^5_\Lambda\mbox{He}$}\\ \hline 
$\Delta$ &$\pm$0.01160&$\pm$0.01507&$\pm$0.04535&0.07258
&0.16800&0.05276\\ 
$\Delta_{zr}$ &$\pm$0.00799&$\pm$0.01038&$\pm$0.03123&0.04998
&0.14414&0.03610\\ 
$\frac{m_{23}}{m_3}\kappa$, fm$^{-1}$ &0.46289&0.46289 
&0.46289&0.46289&0.18325&1.61031\\ 
$1/a_B$, fm$^{-1}$ &0.00370&0.00480&0.01446&0.02314
&0.02641&0.05813\\ 
$p_0^b$, fm$^{-1}$ 
&0.06870&0.07830&0.13582&0.17182&0.11131&0.34356\\ \hline \vspace{5mm}
\end{tabular}
\end{table}

\pagebreak

\noindent {\bf Figure captions} \\ [6mm]

\noindent Fig. 1. The quantity $A$ as a function of the incident momentum
$p_0$ for the deuteron and two lambda hypernuclei, $^3_\Lambda\mbox{H}$ 
and $^5_\Lambda\mbox{He}$, calculated by the formulae (81) and (82). The
solid curves correspond to the separable potential model (9) and (72)
with the parameters (98) -- (100), the dotted curves --- to the zero-range 
interaction ($\beta \rightarrow \infty$). \\ [3mm]

\noindent Fig. 2. The deviation $\Delta$ versus $p_0$ for $ \mu^{+}-\mbox{d}$ 
and $\pi^{+}-\mbox{d}$ scattering calculated by the formulae (81) and (82) that 
correspond to the separable neutron-proton interaction potential (9) and 
(72) with the parameters (98). \\ [3mm]
           
\noindent Fig. 3. The deviation $\Delta$ versus $p_0$ for 
$\mbox{K}^{+}-\mbox{d}$ scattering calculated by the formulae (81) and (82) 
that correspond to the separable neutron-proton interaction potential (9) and 
(72) with the parameters (98). \\ [3mm]

\noindent Fig. 4. The deviation $\Delta$ versus $p_0$ for $\mbox{p}-\mbox{d}$ 
(solid line), $\mbox{p}-^3_\Lambda\mbox{H}$ (dashed line) and
$\mbox{p}-^5_\Lambda\mbox{He}$ (dotted line) scattering calculated by the 
formulae (81) and (82) that correspond to the separable n-p, $\Lambda-\mbox{t}$ 
and $\Lambda-\alpha$ interaction potential (9) and (72) with the parameters 
(98) -- (100).

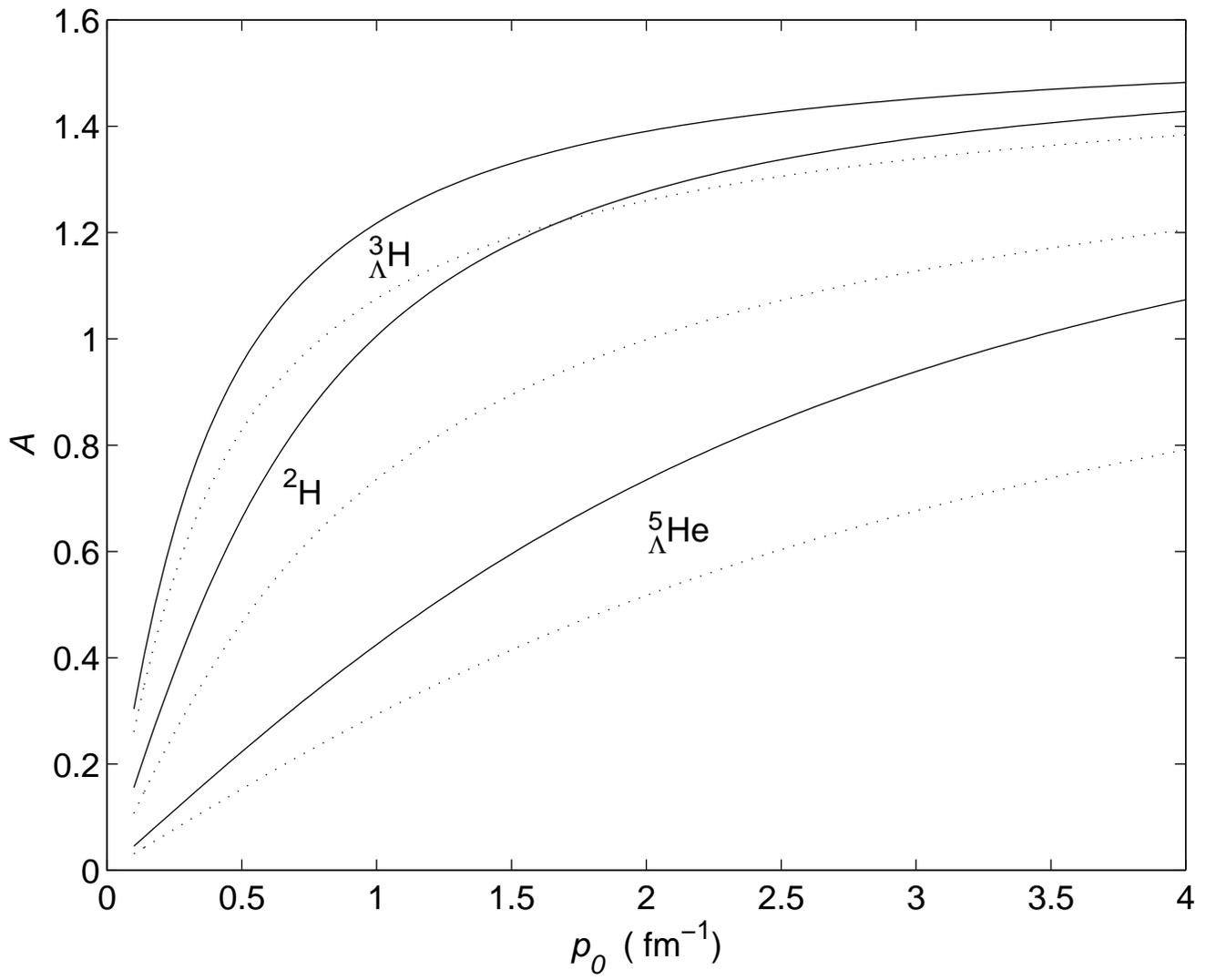
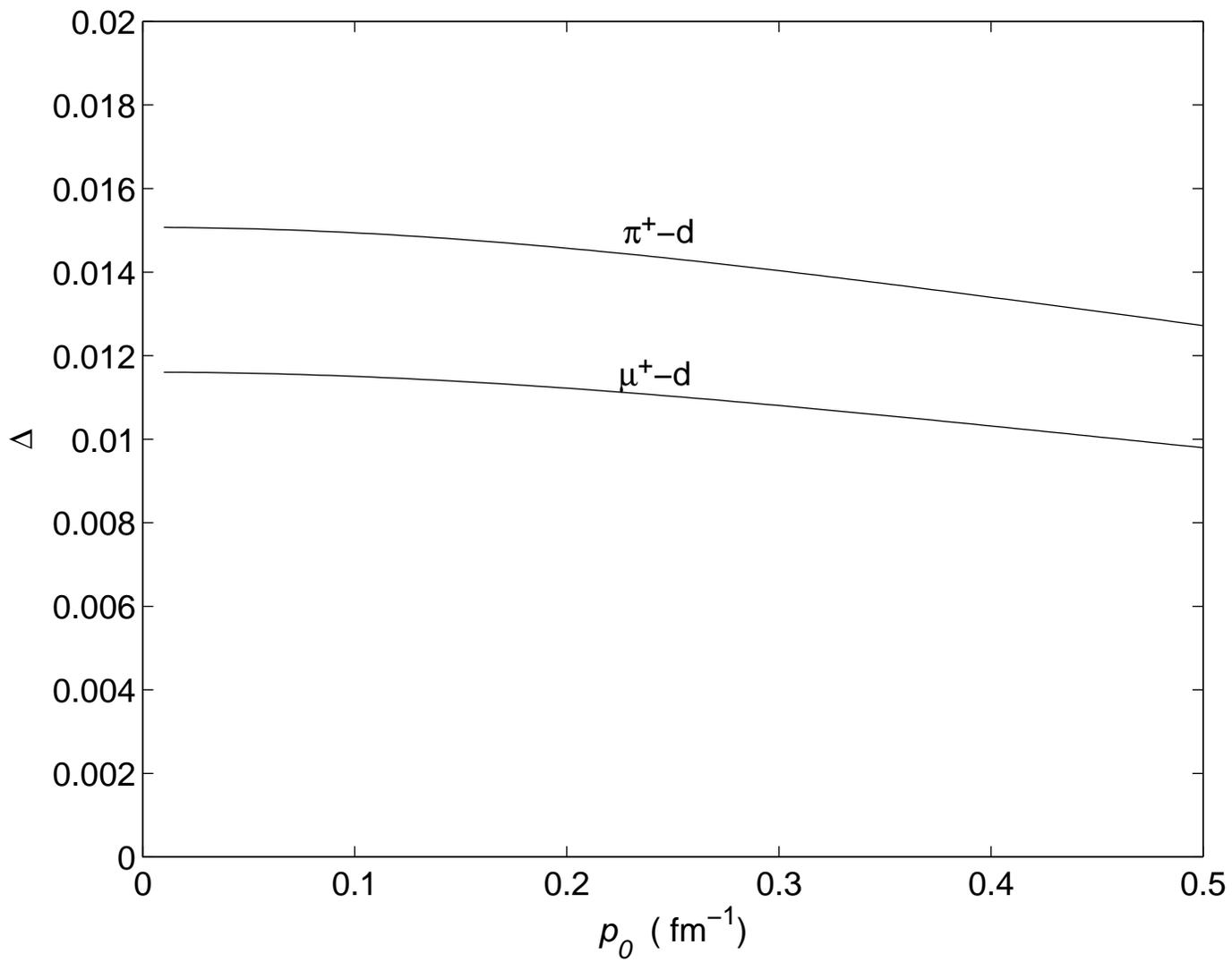
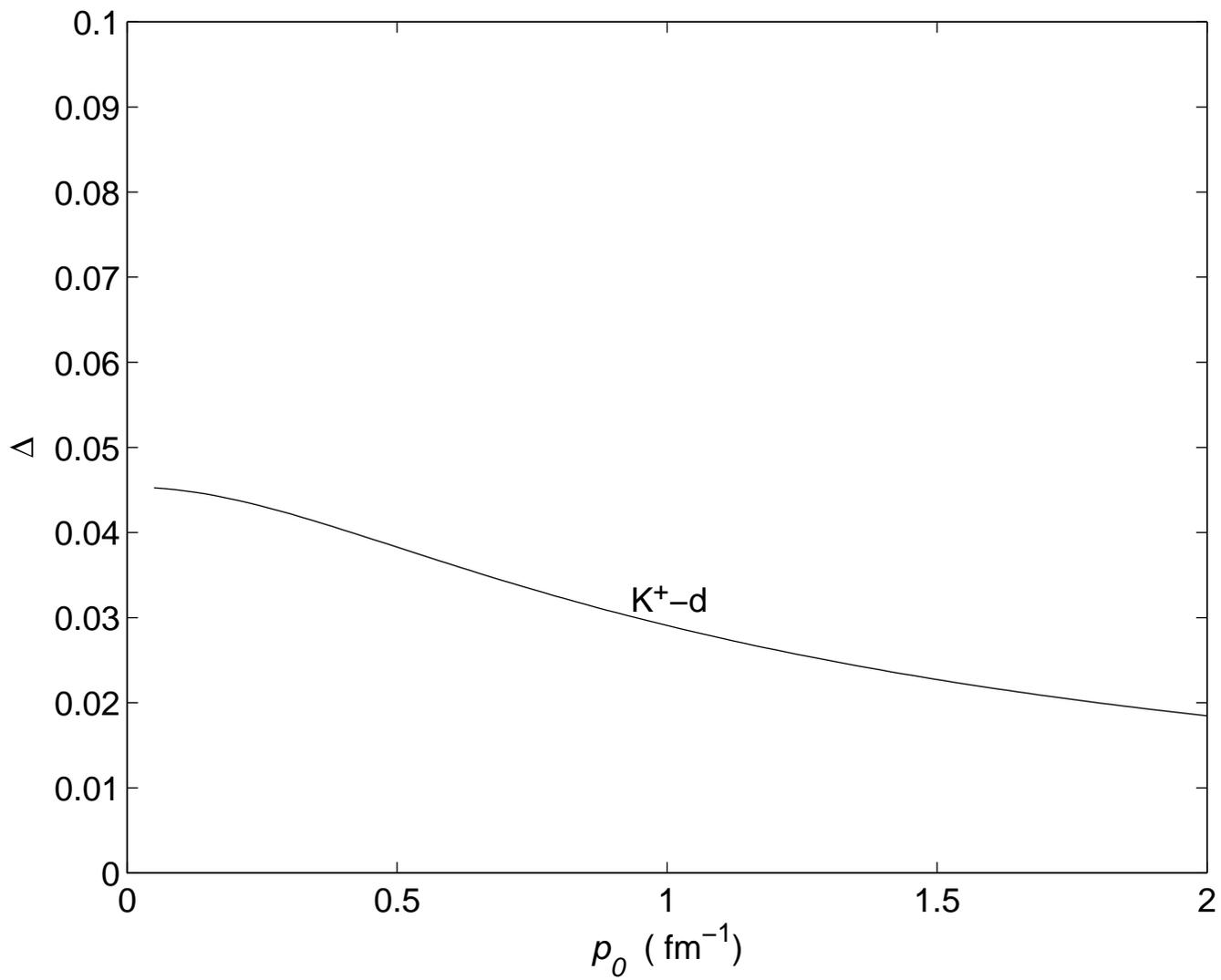
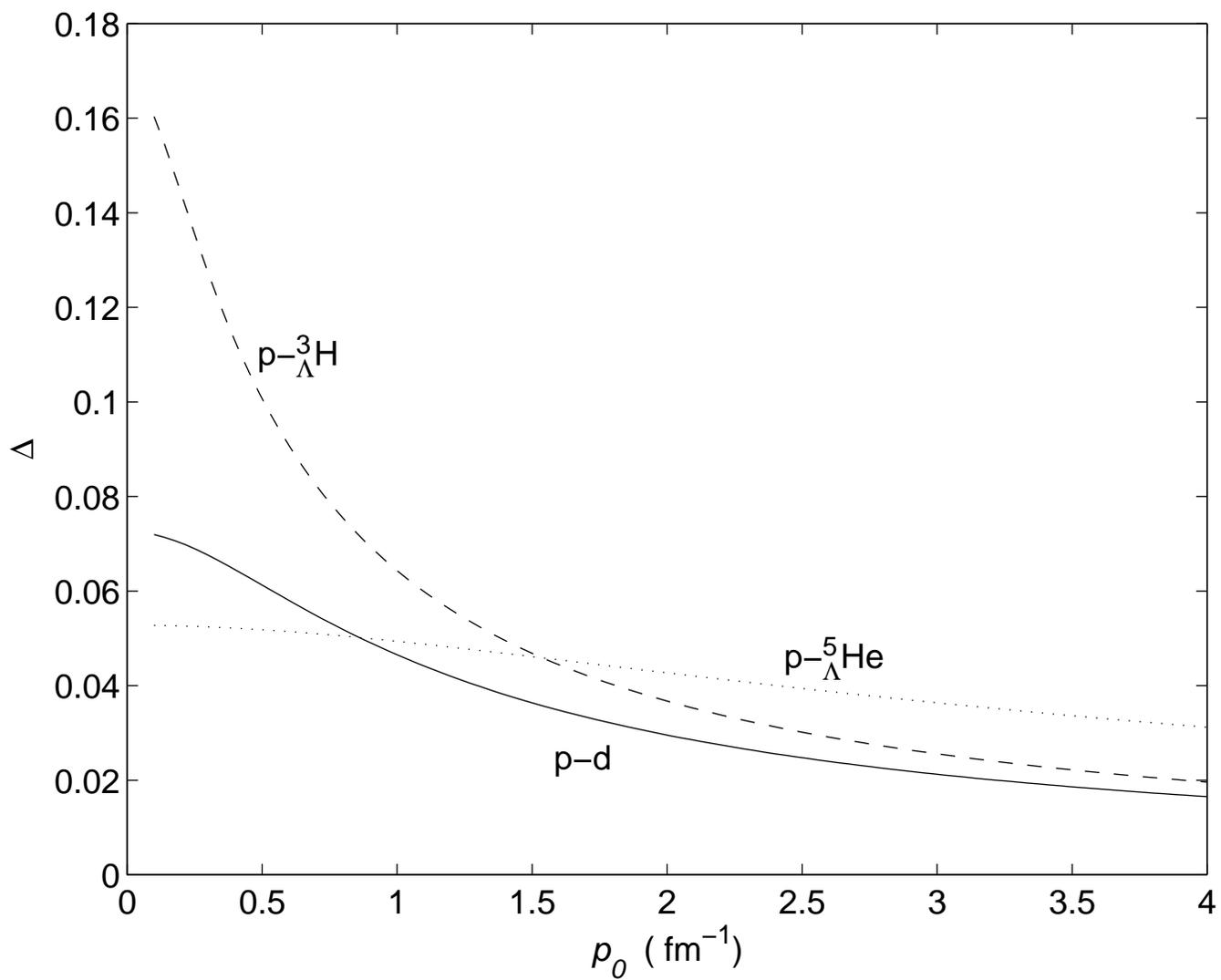
\end{document}